\documentclass[a4paper,fleqn,usenatbib]{mnras}

\usepackage[T1]{fontenc}
\usepackage{ae,aecompl}

\usepackage{graphicx}	
\usepackage{amsmath}	
\usepackage{amssymb}	
\usepackage{array}

\newcommand{\erg}{erg cm$^{-2}$ s$^{-1}$} 
\newcommand{\xm}{\emph{XMM-Newton}}
\newcommand{\sw}{\emph{Swift}/XRT}
\newcommand{\nus}{\emph{NuSTAR}}
\newcommand{\mxb}{MXB 1658--298}

\title[Spectral properties of \mxb]{Spectral properties of \mxb\ in the low/hard and high/soft state}

\author[R. Sharma et al.]{Rahul Sharma$^{1}$\thanks{E-mail: rahul1607kumar@gmail.com},
Abdul Jaleel$^{1}$,
Chetana Jain$^{2}$,
Jeewan C. Pandey$^{3}$,
Biswajit Paul$^{4}$
and \newauthor 
Anjan Dutta$^{1}$
\\
$^{1}$Department of Physics and Astrophysics, University of Delhi, Delhi 110007, India\\
$^{2}$Hansraj College, University of Delhi, Delhi 110007, India\\
$^{3}$Aryabhatta Research Institute of Observational Sciences (ARIES), Manora Peak, Nainital, Uttarakhand 263002, India\\
$^{4}$Raman Research Institute, Sadashivnagar, C. V. Raman Avenue, Bangalore 560080, India\\
}

\date{Accepted XXX. Received YYY; in original form ZZZ}

\pubyear{2018}

\begin{document}
\label{firstpage}
\pagerange{\pageref{firstpage}--\pageref{lastpage}}
\maketitle

\begin{abstract}

We report results from a broadband spectral analysis of the low mass X-ray binary \mxb\ with the \sw\ and \nus\ observations made during its  2015--2017 outburst. The source showed different spectral states and accretion rates during this outburst. 
The source was in low/hard state during 2015; and it was in high/soft state during the 2016 \nus\ observations.
This is the first time that a comparison of the soft and hard spectral states during an outburst is being reported in \mxb. Compared with the observation of 2015, the X-ray luminosity was about four times higher during 2016. 
The hard state spectrum can be well described with models consisting of a single-temperature blackbody component along with Comptonized disc emission, or three component model comprising of multi-colour disc, single-temperature blackbody and thermal Comptonization components. 
The soft state spectrum can be described with blackbody or disc blackbody and Comptonization component, where the neutron star surface (or boundary layer) is the dominant source for the Comptonization seed photons. We have also found a link between the spectral state of the source and the Fe K absorber. The absorption features due to highly ionized Fe were observed only in the soft state. This suggests a probable connection between accretion disc wind (or atmosphere) and spectral state (or accretion state) of \mxb.  

\end{abstract}




\section{Introduction}

Low Mass X-ray Binaries (LMXBs) consist of a neutron star (NS) or a black hole (BH) with a low mass ($\lesssim 1 M_{\sun}$) companion orbiting around it. The compact object accretes matter from its companion and the accretion rate is usually inferred from the position of the sources in their X-ray colour-colour diagrams (CDs) or hardness-intensity diagrams (HIDs) \citep{2006book}. The low-luminosity sources trace out an atoll-like shape, with the different branches referred to as the banana branch (high inferred accretion rate) and the island state (low inferred accretion rate) \citep{Hasinger}. They have luminosity in the range of 0.001--0.5 $L_{Edd}$ \citep{vanderKlis}. 

The NS LMXBs display spectral states and hysteresis pattern similar to that observed in BH binaries, indicating similarity between these two kinds of sources \citep{Munoz}. NS spectral emission is composed of three main components: disc blackbody from the accretion disc and Comptonization component due to Comptonization of soft X-rays in the corona as of the BH spectra \citep{Done}, with the addition of a blackbody component for emission from the NS surface or boundary layer \citep{Lin07, Lin09,Armas}. During the soft state, the emission is usually dominated by the thermal components, with characteristic temperatures of $kT \sim 0.5-2.0$ keV \citep{Bloser2000, Sakurai2012, Armas}.  
The Comptonized component is weak with low temperature and large optical depth. In the hard state, the spectra are dominated by a hard Comptonized component with temperatures of a few tens of keV and low optical depths. The thermal components are observed at low temperatures ($kT < 1$ keV) and with significantly lower luminosity \citep{Raichur2011, Sakurai2012, Zhang2016,Armas}.

Narrow absorption features from highly ionized Fe and other elements have been observed in large number of X-ray binaries. In particular, Fe \textsc{xxv} (He-like) or Fe \textsc{xxvi} (H-like) absorption lines near 6.6--7.0 keV have been reported from several NS LMXBs \cite[e.g.,][]{Sidoli, Trigo, Hyodo2009, Ponti14, PontiAX-2015, Raman2018} that are almost all viewed at fairly high inclination angles; most of them being dippers \citep{Boirin2004,Boirin2005,Trigo}. This indicates that the highly ionized plasma probably originates in an accretion disc atmosphere or wind, which could be a common feature of accreting binaries but primarily detected in systems viewed close to edge-on \citep{Trigo2013,Trigo2016}. 

\mxb\ is a transient atoll NS LMXB system \citep{Wijnandsbb} which shows eclipses, dips and thermonuclear bursts in its X-ray lightcurves.
It was discovered in 1976 with SAS-3 \citep{Lewin1976}. It has an orbital period of $\sim 7.1$ h and its eclipse lasts for $\sim 15$ min \citep{CominskynWood}. This active phase lasted for about 2 years. 
The second outburst was observed in 1999--2001 \citep{intZand} during which coherent burst oscillations at frequency of $\sim 567$ Hz were reported \citep{WijnandsBO}. The 0.5--30 keV \emph{BeppoSAX} spectrum obtained from observations made during this outburst, was modeled by a combination of a soft disc blackbody and a harder Comptonized component \citep{Oosterbroek}. The residuals to this fit suggested the presence of emission features due to Ne-K/Fe-L and Fe-K in the spectrum. The observations made with \xm\ revealed the presence of narrow resonant absorption features due to O \textsc{viii}, Ne \textsc{x}, Fe \textsc{xxv} and Fe \textsc{xxvi}, together with a broad Fe emission feature \citep{Sidoli}. 
The third outburst from the source was detected in the year 2015 \citep{Negoro,Bahramian}, which lasted for $\sim 1.5$ years \citep{Parikh}. 
The evolution of the orbital period over about 40 years of data indicates the presence of a circumbinary planet around this binary system \citep{Jain2017}.

In this work, we report results of spectral study of the persistent emission of \mxb\ by using the \sw\ and \nus\ observations made during 2015 and 2016. We also report the detection of state dependent absorption features due to highly ionized materials. 

\section{Observations and data reduction}

The Nuclear Spectroscopic Telescope ARray \citep[\nus;][]{Harrison} mission features two telescopes, focusing X-rays between 3 and 79 keV onto two identical focal planes (usually called focal plane modules A and B, or FPMA and FPMB). During its 2015--2017 outburst, \mxb\ was observed twice with \nus. The observation details are given in Table \ref{obslog}. We have used the most recent \nus\ analysis software distributed with HEASOFT version 6.20 and the latest calibration files (version 20170120) for reduction and analysis of the \nus\ data. The calibrated and screened event files have been generated by using the task \textsc{nupipeline}.
A circular region of radius $100''$ centered at the source position was used to extract the source events. Background events were extracted from a circular region of same size away from the source. The task \textsc{nuproduct} was used to generate the lightcurves, spectra and response files. Using \textsc{grppha}, the spectra were grouped to give a minimum of 200 counts per bin. The FPMA/FPMB light curves were background-corrected and summed using \textsc{lcmath}. The persistent spectra were extracted by excluding the dips, eclipses and bursts.

\begin{table}
\caption{Log of X-ray observations.}
\centering
\resizebox{\columnwidth}{!}{
\begin{tabular}{c c c c c}
\hline \hline
Satellite & OBS Id & Start Time & Exposure$^*$ & Mode\\
&& (Date hh:mm:ss)&(ks)\\\hline
\nus\ & 90101013002 & 2015-09-28 21:51:08& 49.7 &-\\
\sw\ & 00081770001 & 2015-09-28 21:36:43& 1.3 & PC\\
\hline
\nus\ & 90201017002 & 2016-04-21 14:41:08& 21.6 &-\\
\sw\ & 00081918001 & 2016-04-21 20:39:01& 0.67 & WT\\
\hline
\multicolumn{5}{l}{$^*$Net exposure time of persistent emission. }\\
\end{tabular}}
\label{obslog}
\end{table}

\mxb\ was also monitored by Neil Gehrels \emph{Swift} Observatory \citep{Gehrels} during its 2015--2017 outburst. For this work, we have focused only on those observations (listed in Table \ref{obslog}) which coincided with the \nus\ observations.
The \emph{Swift}/X-ray telescope \citep[XRT;][]{Burrows} data were analysed using standard tools incorporated in HEASOFT version 6.20.
The 2015 Swift/XRT observation was obtained in the PC mode wherein the effects of photon pile-up have been corrected\footnote{http://www.swift.ac.uk/analysis/xrt/pileup.php}. The 2016 observation was obtained in the WT mode and it was not affected by pile-up as the XRT count rate was $<100$ count/sec  \citep{Romano}. We have used \textsc{xselect} to extract source events from a circular region of $70.8''$(=30 pixel) radius for WT mode; and annulus region of inner radius $10''$ and outer radius $100''$ for PC mode. Background events were obtained from the outer regions of the CCD. 
For the WT mode, we have extracted background from a region with same size of the source and for the PC mode, circular region of radius $150''$ was used. Exposure maps were used to create ancillary response files with \textsc{xrtmkarf}, to account for hot pixels and bad columns on the CCD. The latest response matrix file was sourced from the calibration data base (version 20160609). Spectra were grouped to contain a minimum of 20 counts per bin.

We have used XSPEC \citep{Arnaud} version 12.9.1 for the spectral fitting. The persistent spectra extracted from \sw, \nus-FPMA and \nus-FPMB observations were fitted simultaneously. We have added a constant to account for cross-calibration of different instruments. The value of constant for \nus-FPMA was fixed to 1 and set free for others.
Due to the low energy spectral residuals in the WT mode, the WT mode data was fitted in the energy range 0.5--10 keV\footnote{http://www.swift.ac.uk/analysis/xrt/digest\_cal.php}; and 0.3--10 keV energy range was used for PC mode data. The \nus\ data in 3--70 keV and 3--35 keV energy range were used for spectral fitting for 2015 and 2016 observations, respectively.
 The photo-electric absorption cross section of \citet{Verner} and abundance of \citet{Wilms} have been used throughout. All the spectral uncertainties and the upper limits reported in this paper are at 90\% confidence level. We have assumed the source distance to be equal to $10$ kpc \citep{Muno,Sharma}.

\begin{figure*}
\centering

\includegraphics[width=0.85\columnwidth]{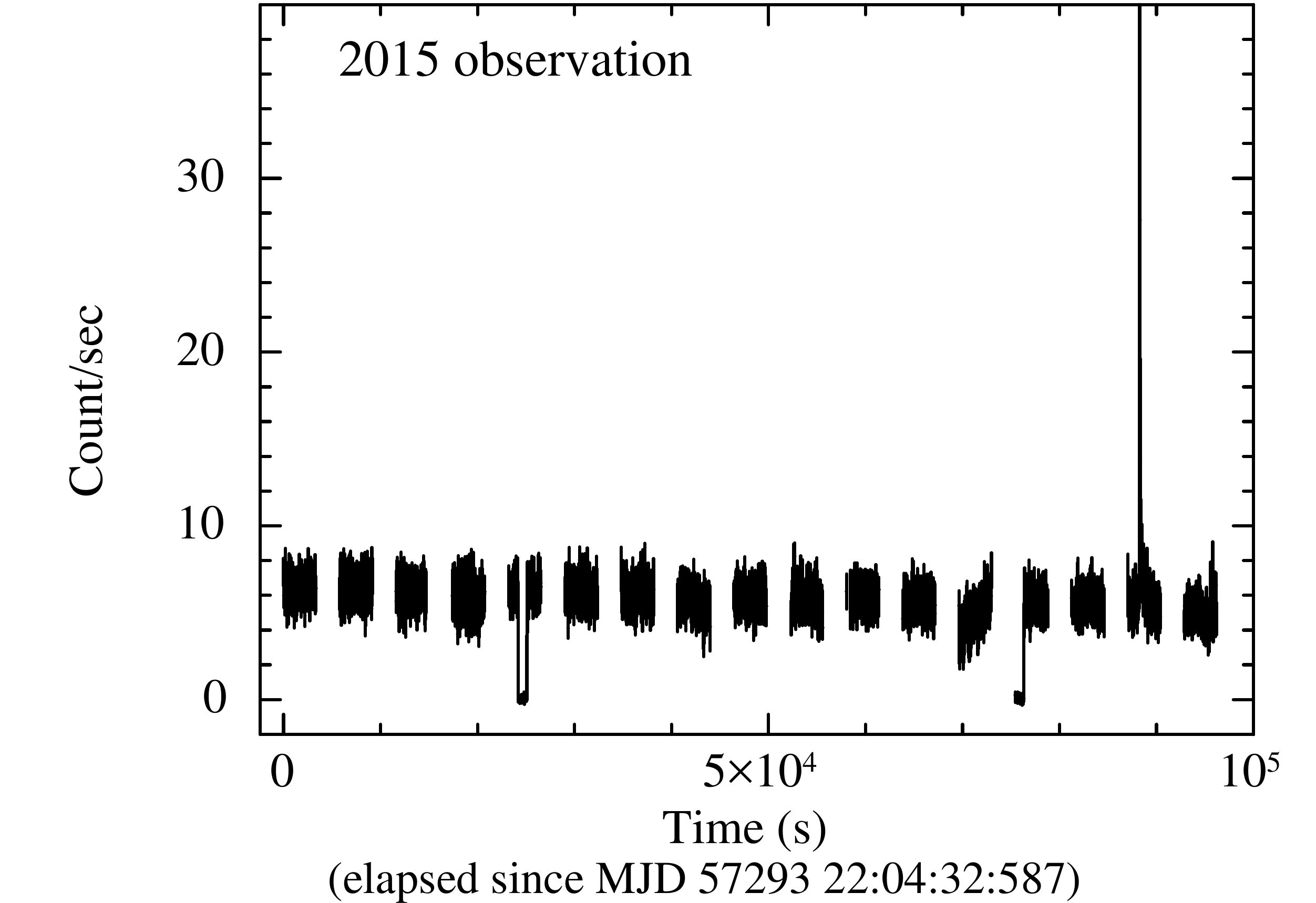}\hspace{1cm}
\includegraphics[width=0.82\columnwidth]{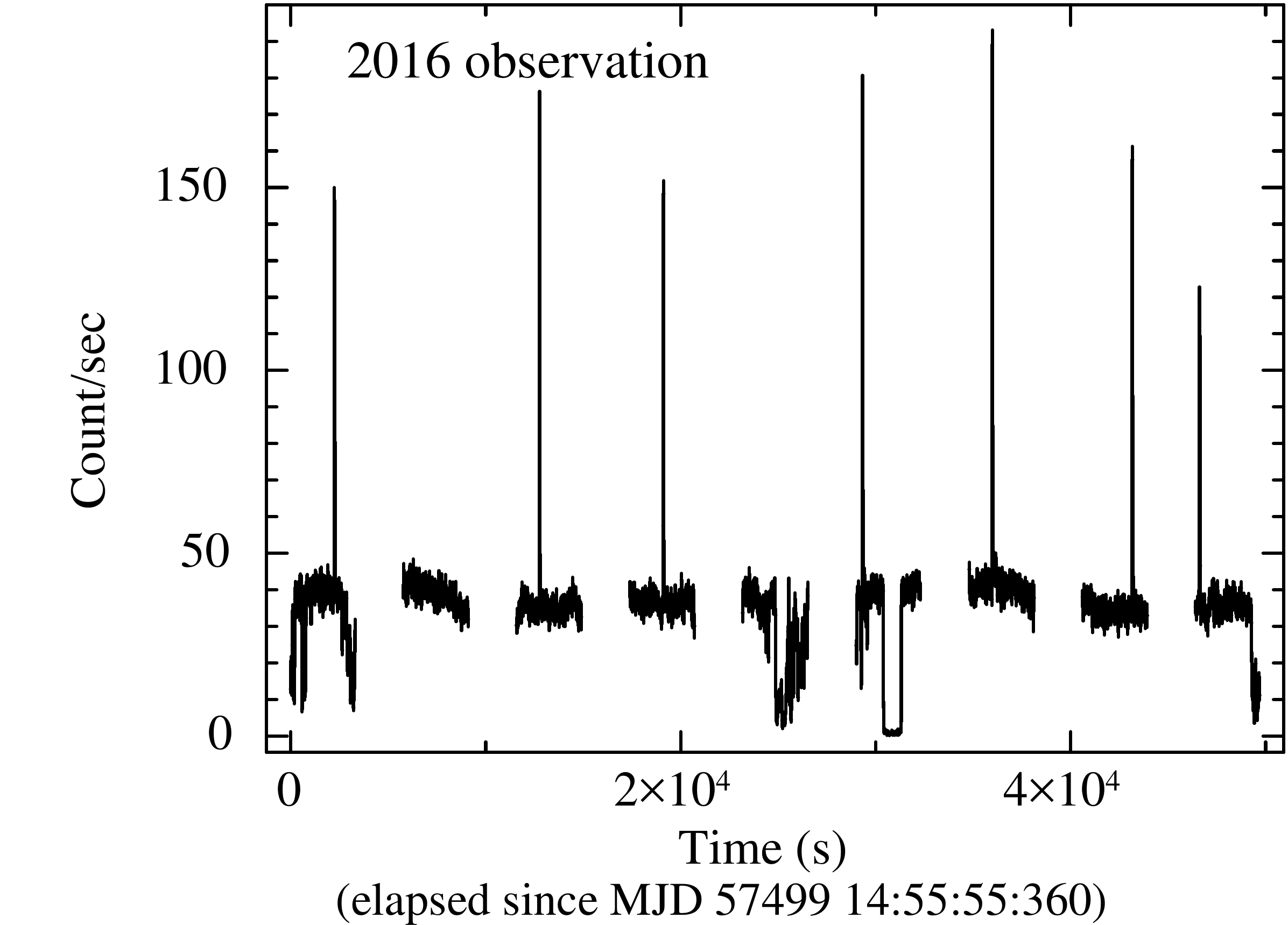}
\caption{ Background subtracted lightcurve of \mxb\ extracted from 2015 (left) and 2016 (right) \nus\ observations binned at 20 sec in energy band of 3--78 keV. }
\label{lc}
\end{figure*}

\section{Lightcurve \& Spectral States}

Figure \ref{lc} shows the background subtracted FPMA+FPMB lightcurves of \mxb\ binned at 20 sec in energy range of 3--78 keV extracted from \nus\ observations of 2015 and 2016. During the 2015 observation, the background subtracted average count rate during the persistent phase was $\sim 6$ count/sec. One type-I thermonuclear X-ray burst was also observed where burst count rate reached $>500$ count/s \citep{Sharma}. During 2016, the background subtracted average persistent count rate increased to $\sim 40$ count/sec and the bursts observed were weaker and more frequent than 2015 observation.

\begin{figure*}
\centering
\includegraphics[width=0.82\columnwidth]{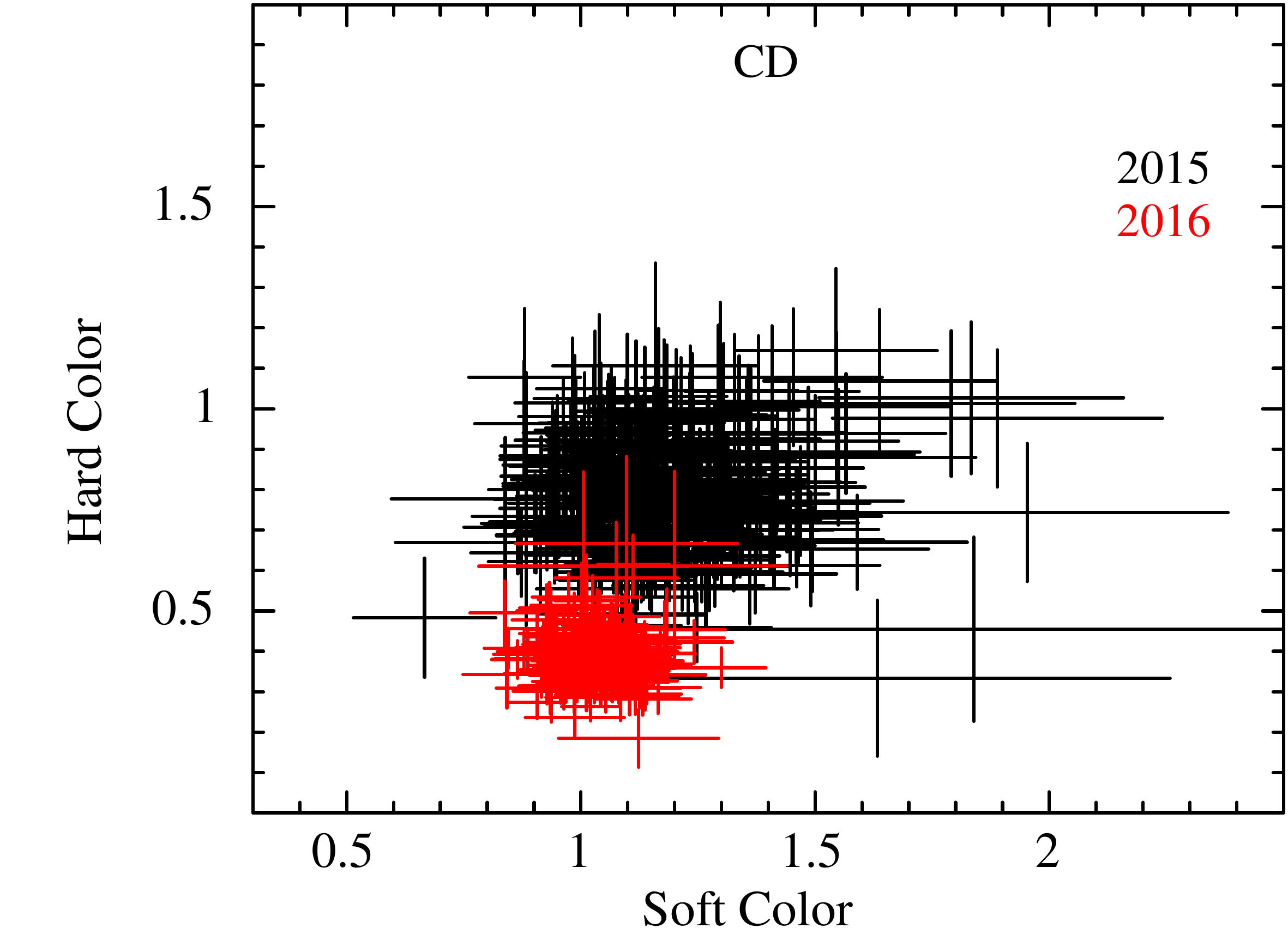}\hspace{1.1cm}
\includegraphics[width=0.82\columnwidth]{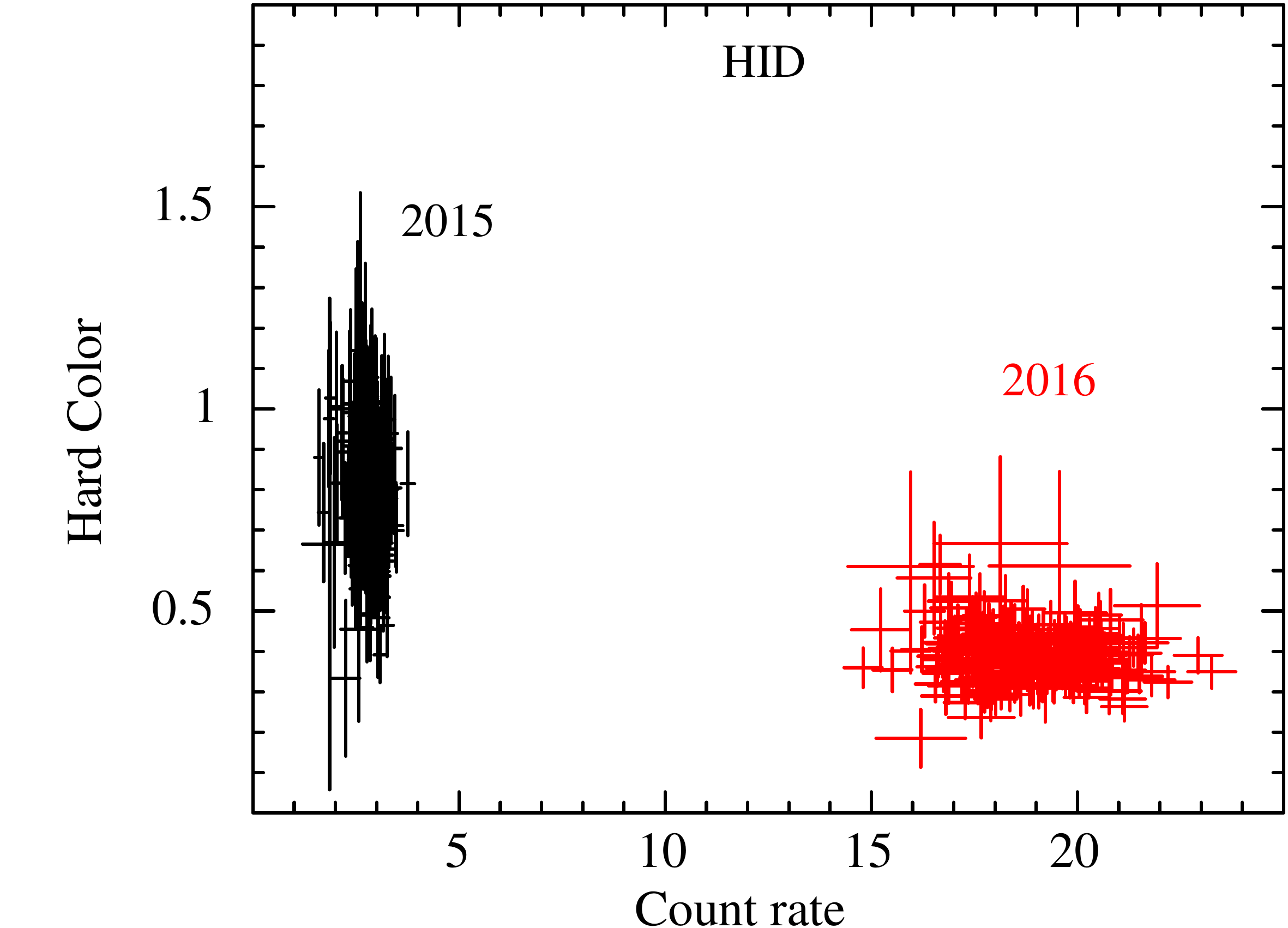}
\caption{\textit{(left)} Colour-colour diagram and \textit{(right)} Hardness-intensity diagram of \mxb. The time resolution of the data is 200 sec and the count rate is in the energy range of 3--30 keV.}
\label{color}
\end{figure*}

\begin{figure}
\centering
\includegraphics[width=0.9\columnwidth]{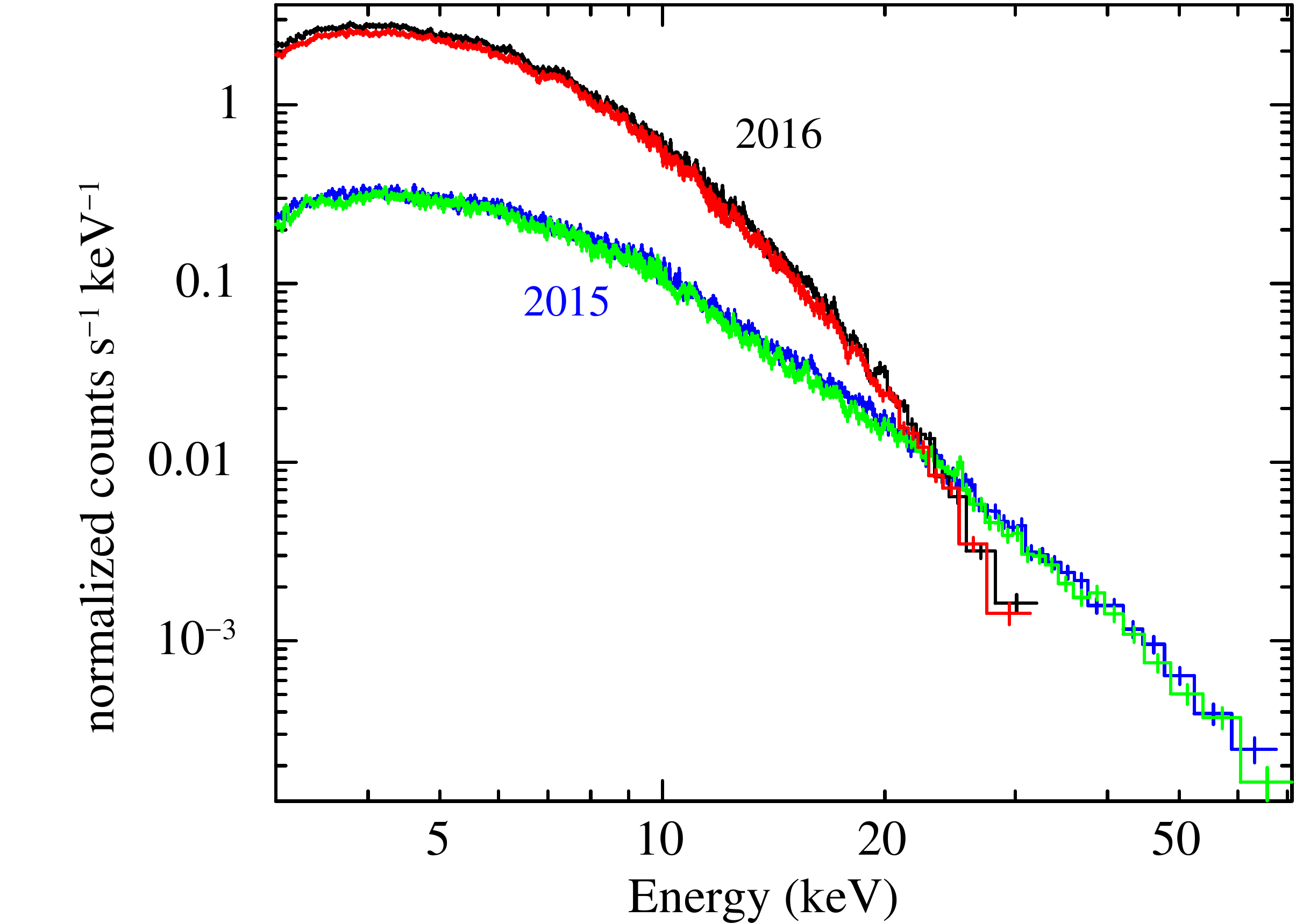}
\caption{Persistent spectrum extracted from \nus\ (FPMA \& FPMB) observations of 2015 and 2016. The spectrum of 2015 was hard and had low flux emission while spectrum of 2016 was soft and had high flux emission. (The black and red colour represent the FPMA and FPMB data of 2016 observation, and blue and green represent FPMA and FPMB data of 2015 observation.)}
\label{compare}
\end{figure}

The CD and HID for \mxb\ are shown in Figure \ref{color}. They have been created after removing data during the bursts, dips and the eclipses. The soft colour corresponds to the ratio of count rate between 5--8 keV and 3--5 keV and the hard colour is the ratio between 12--30 keV and 8--12 keV. In Figure \ref{color}, the data points of 2015 and 2016 observations are denoted by black and red colours, respectively. In this figure, two branches are visible; the lower branch (red) can be identified with the banana branch and the upper (black) with the island state in CD (left panel of Figure \ref{color}). These different spectral branches depend on the accretion rates \citep{Hasinger}. Observation of 2015 shows low accretion state (island) and 2016 is showing high accretion state (banana), as shown in HID (right panel of Figure \ref{color}). For a quick comparison, the persistent spectrum for both observations is shown in Figure \ref{compare}. It clearly shows that the spectrum of 2015 was low/hard and 2016 was high/soft.

\section{Spectral Analysis and Results}


\begin{table*}
\centering
\caption{Spectral parameters of MXB 1658-298 obtained from the different best fit models in low/hard state.}
\resizebox{0.75\linewidth}{!}{
\begin{tabular}{c c c c c c}
\\\hline
Component 	& parameters 			& Model A1 		& Model A2		& Model A3		& Model A4	\\  [0.5ex]
\hline
\textsc{TBabs}	& $N_H ( 10^{22}~cm^{-2})$	& $0.25 \pm 0.07$ & $<0.18$ & $0.23^{+0.07}_{-0.06}$& $0.23^{+0.08}_{-0.06}$ \\  [0.5ex]
\hline
\textsc{diskbb} & $kT_{disc}~(keV)$		& &			& $0.63^{+0.10}_{-0.08}$& $0.56^{+0.13}_{-0.12}$ \\  [0.5ex]
		& $Norm_{disc}$			& &			& $15.5^{+13.7}_{-7.4}$	& $10^{+18}_{-6}$ \\  [0.5ex]
\hline
\textsc{bbodyrad}& $kT_{BB}~(keV)$		& $1.04 \pm 0.04$ & $1.04 \pm 0.03$ & $1.07^{+0.10}_{-0.08}$ & $1.14^{+0.09}_{-0.08}$ \\  [0.5ex]
		& $Norm_{BB}$			& $1.13 \pm 0.17$ & $1.13^{+0.17}_{-0.15}$ & $1.46^{+0.58}_{-0.46}$ & $0.80^{+0.27}_{-0.23}$ \\  [0.5ex]
\hline
\textsc{nthcomp}& $\Gamma$			& $1.74^{+0.01}_{-0.02}$	&& $1.71 \pm 0.03$	& $1.71 \pm 0.03$	\\  [0.5ex]
		& $kT_{e}~(keV)$		& $19.2^{+3.8}_{-2.3}$	&& $17.1^{+3.0}_{-2.0}$	& $16.8^{+2.8}_{-1.9}$	\\  [0.5ex]
		& $kT_{seed}~(keV)$		& $0.24 ~(<0.45)$	&& $=kT_{BB}$		& $=kT_{disc}$		\\  [0.5ex]
		& input\_type			& 1 			&& 0			& 1			\\  [0.5ex]
		& Norm ($\times 10^{-2}$)	& $1.6^{+0.2}_{-0.3}$	&& $0.13 \pm 0.13$& $1.05^{+0.22}_{-0.18}$\\  [0.5ex]
\hline
\textsc{compTT} & $kT_{e}~(keV)$		&& $14.8^{+1.6}_{-1.2}$ &\\ [0.5ex]
		& $kT_{seed}~(keV)$		&& $0.28^{+0.05}_{-0.07}$ &\\ [0.5ex]
		& $\tau$			&& $6.0 \pm 0.4$ &\\ [0.5ex]
		& Norm ($\times 10^{-3}$)	&& $2.1^{+0.5}_{-0.3}$ &\\ [0.5ex]

\hline
		& $^afrac_{disc}$     & - & - & $0.20$ & $0.08$ \\  [0.5ex]
		& $^afrac_{BB}$        & $0.05$ & $0.06$ & $0.08$ & $0.06$  \\  [0.5ex]
		& $^afrac_{comp}$      & $0.95$ & $0.94$ & $0.72$ & $0.86$  \\  [0.5ex]
\hline
		&$^{\dagger}$Flux$_{0.5-10~keV}$	& $1.15^{+0.05}_{-0.04}$ & $1.06 \pm 0.04$ & $1.18 \pm 0.01$ & $ 1.18 \pm 0.02 $ \\  [0.5ex]
		&$^{\dagger}$Flux$_{0.1-100~keV}$  & $2.63^{+0.24}_{-0.08}$ & $2.40^{+0.08}_{-0.07}$ & $2.61 \pm 0.05$ & $ 2.61 \pm 0.05 $ \\  [0.5ex]
		&$^{\dagger\dagger}L_X$         & $3.15^{+0.29}_{-0.10}$ & $2.87^{+0.09}_{-0.08}$ & $3.12 \pm 0.06$ & $ 3.12 \pm 0.06 $ \\  [0.5ex]
\hline
		&$\chi^2/dof$			& 712.4/643	& 709/643	&703.1/642	& 703.4/642		\\  [0.5ex]
\hline

\multicolumn{6}{l}{Model A1 : \texttt{TBabs$\times$(bbodyrad+nthcomp[diskbb])};}\\
\multicolumn{6}{l}{Model A2 : \texttt{TBabs$\times$(bbodyrad+compTT)};}\\
\multicolumn{6}{l}{Model A3 : \texttt{TBabs$\times$(diskbb+bbodyrad+nthcomp[BB])};}\\
\multicolumn{6}{l}{Model A4 : \texttt{TBabs$\times$(diskbb+bbodyrad+nthcomp[diskbb])};}\\
\multicolumn{6}{l}{$^\dagger$Unabsorbed flux in units of $10^{-10}$ \erg.}\\
\multicolumn{6}{l}{$^{\dagger\dagger}$Unabsorbed 0.1--100 keV X-ray Luminosity in units of $10^{36}$ erg s$^{-1}$.}\\
\multicolumn{6}{l}{$^a$represents the component fraction.}\\
\end{tabular}}
\label{table-para2015}
\end{table*}

\begin{figure*}
\includegraphics[width=0.69\columnwidth]{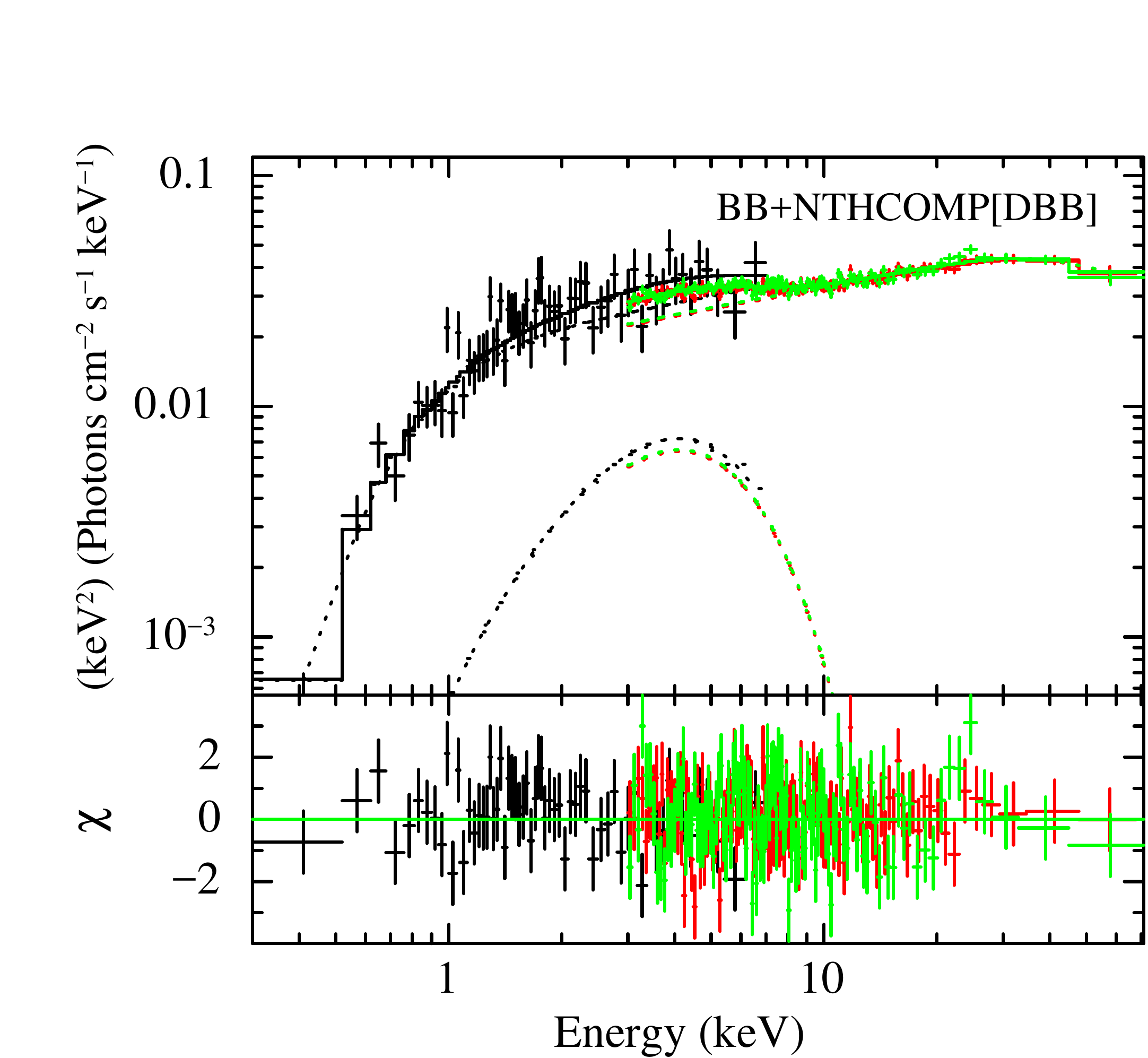}
\includegraphics[width=0.69\columnwidth]{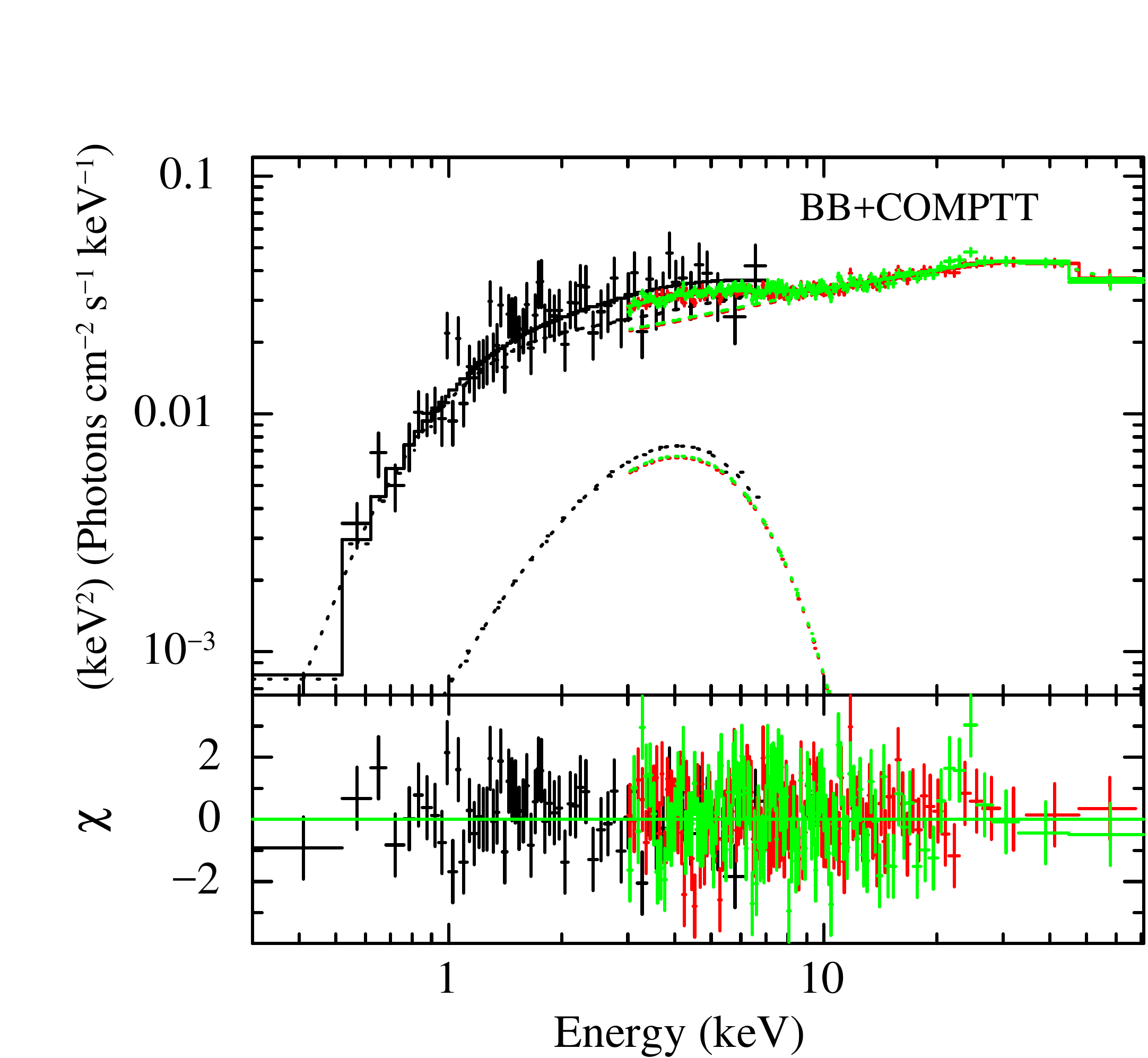}

\includegraphics[width=0.69\columnwidth]{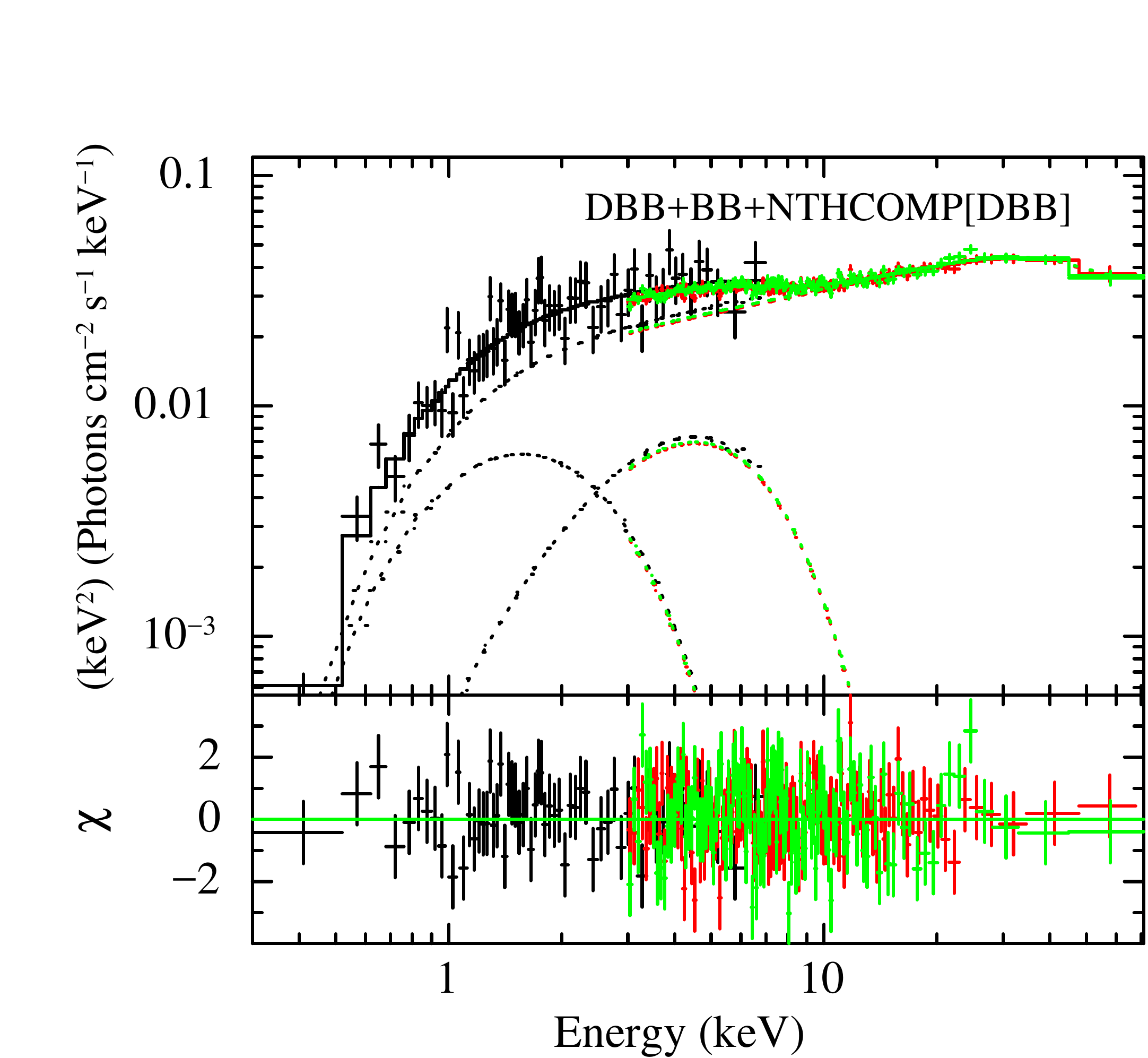}
\includegraphics[width=0.69\columnwidth]{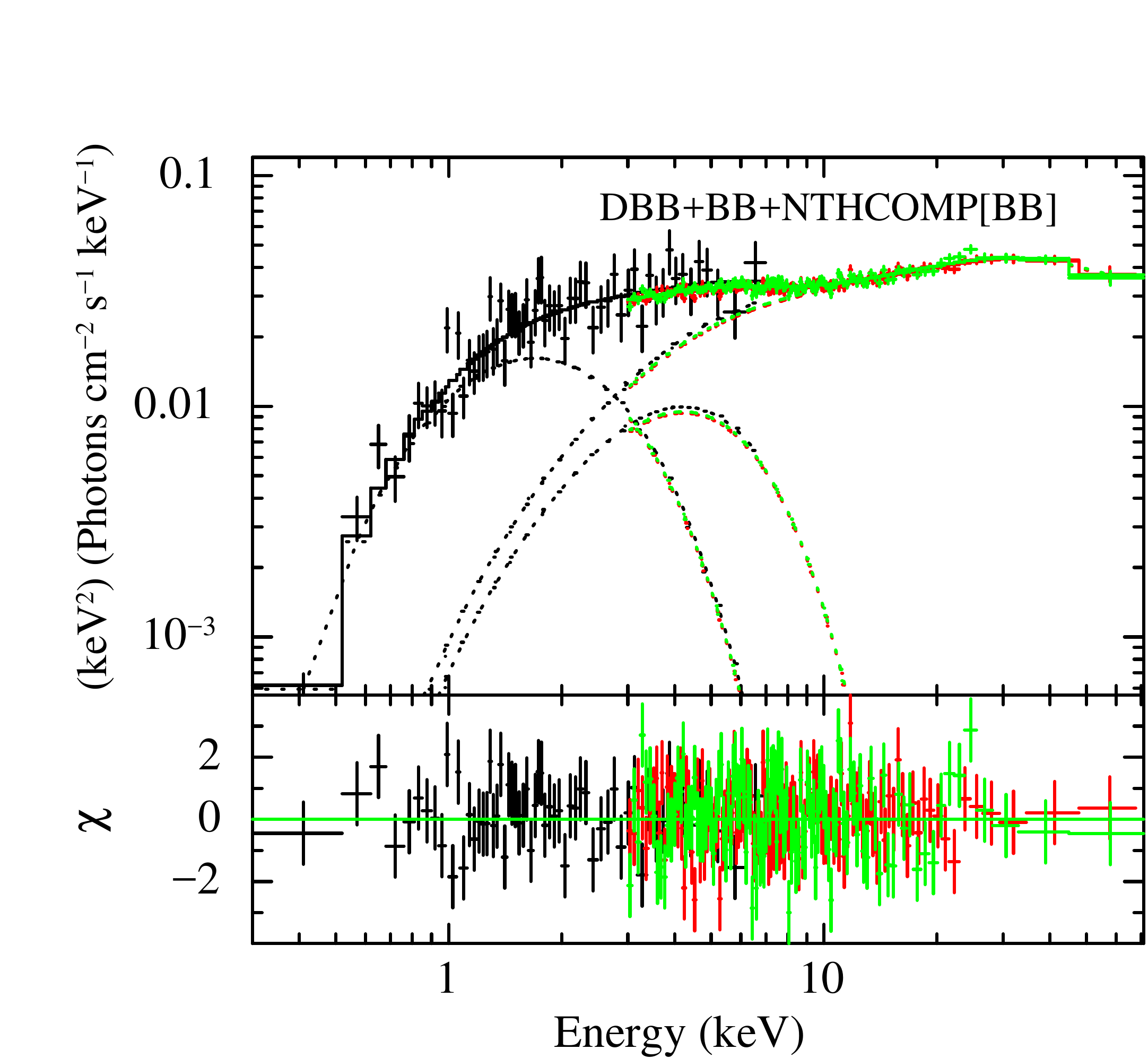}

\caption{ Spectrum of MXB 1658-298 extracted from 2015 observation (hard state) fitted with combinations of thermal and non-thermal components (model is mentioned on the top-right of each figure; DBB--diskbb, BB--bbodyrad). The \nus\ data have been rebinned for representation purpose. The black points represent the \sw, red represent \nus-FPMA and green represent \nus-FPMB spectra.}
\label{fig-pe2015}
\end{figure*}


The persistent emission spectra of \mxb\ have been modeled with two component models \citep{Oosterbroek, Sidoli, Trigo, Sharma}. During the 1999--2001 outburst, \mxb's broadband spectrum was studied with \emph{BeppoSAX} and was modeled with combination of disc blackbody and Comptonization component \citep{Oosterbroek}. For the same outburst, the spectrum obtained from the observations made with \xm\ was modeled with combination of blackbody and cutoff power-law \citep{Sidoli}. 
In this work, we have modeled the persistent emission spectrum obtained from the observation of 2015 (low/hard) and 2016 (high/soft) with different combinations of continuum. This kind of comprehensive spectral analysis is being done for the first time in \mxb. We have tried single component models for the fitting the spectra of low/hard and high/soft states. But they do not provide statistically adequate fit in both the cases.

\subsection{The Hard State Emission}
\label{hardstate}

We started with single-temperature blackbody (\texttt{bbodyrad}) plus exponential cutoff power-law (\texttt{cutoffpl}) to model the spectrum of 2015. \texttt{TBabs} has been used for modeling the interstellar absorption. The model \texttt{TBabs $\times$ (bbodyrad+cutoffpl)} did not give a good fit, $\chi^2=735$ for 644 degrees of freedom (dof).

Then, we replaced the \texttt{cutoffpl} component by the thermal Comptonization model \texttt{nthcomp} \citep{Zdziarski,Zycki}. We have expressed the Comptonized model as \texttt{nthcomp[BB]} or \texttt{nthcomp[diskbb]}, where \texttt{[BB]} means that seed photons are provided by blackbody and \texttt{[diskbb]} means that seed photons are provided by disc blackbody.
A model comprising of the disc blackbody plus Comptonized blackbody gave unphysical value of inner disc radius and seed photon temperature ($kT_{seed}$). The Comptonized disc plus blackbody component (\texttt{TBabs$\times$(bbodyrad+nthcomp[diskbb])}) provided statistically good fit, $\chi^2/dof=712.4/643$ (model A1, see Table \ref{table-para2015}), where disc emission is completely Comptonized with $kT_{disc} = kT_{seed}=0.24 (<0.45)$ keV. In model A1, \texttt{nthcomp} component specifies the Comptonization via coronal electron temperature $kT_e=19.2^{+3.8}_{-2.3}$ keV and the spectral slope $\Gamma=1.74^{+0.01}_{-0.02}$. Using the equation, 
\begin{equation*}
\centering
\Gamma = -{1 \over 2} + \sqrt{ {9 \over 4} + { 1 \over {kT_e \over m_ec^2} (1 + {\tau \over 3}) \tau }}
\end{equation*}
\citep[see,][]{Zdziarski} optical depth of Comptonizing corona estimated to $\tau \sim 4.1$. Following \citet{intZand1999,Iaria2016}, we have estimated the emission radius of seed photons (assuming a spherical emission) using
 \begin{equation*}
R_0 = 3 \times 10^4 D_{kpc} [ f_{bol}/(1+y) ]^{1/2} (kT_{seed})^{-2} ~km,
\end{equation*}
where $D_{kpc}$ is the source distance in units of kpc, $f_{bol}$ is unabsorbed bolometric Comptonization flux and $y-$parameter is the Compton parameter defined as $y = 4 kT_e \tau^2/m_e c^2$ (relative energy gained by the photons in the inverse Compton scattering). Using best fit values, we obtained $y-$parameter $\sim 2.5$ and $R_0 \sim 44 (>12.5)$ km, where we used $f_{bol}$ to be 0.1--100 keV unabsorbed flux of Comptonized component ($=2.5 \times 10^{-10}$ \erg). The emission radius of seed photons is consistent with the seed photons being generated at the inner accretion disc. The $\sim 1$ keV blackbody with emission radius of $1.06 \pm 0.08$ km suggests the emission from equatorial-belt like region of NS \citep{Bloser2000, Lin07, Sakurai2012, Zhang2014}.  

We also examined the continuum with another thermal Comptonization model \texttt{CompTT} \citep{Titarchuk} along with the same thermal components. \texttt{CompTT} model contains as free parameters; the temperature of the Comptonizing electron $kT_e$, the plasma optical depth $\tau$ and the input temperature of the soft photon (Wein) distribution $kT_{seed}$. A spherical geometry was assumed for the Comptonizing region (corona). The model \texttt{TBabs $\times$ (bbodyrad + CompTT)} provided us a marginally better fit, $\chi^2/dof=709/643$ (model A2, see Table \ref{table-para2015}).

We finally fit the spectrum with the three components model, composed by two thermal components (\texttt{bbodyrad} and \texttt{diskbb}) and Comptonization component, \texttt{nthcomp[BB]} (model A3) or \texttt{nthcomp[diskbb]} (model A4) \citep{Lin07,Armas}. We obtained comparable chi-square values in the both cases.
The best fit parameters obtained from the above described models are reported in Table \ref{table-para2015}. Figure \ref{fig-pe2015} shows the best fit spectrum of \mxb\ in the hard state from all the four best fit models explained above. We used the \texttt{cflux} convolution model in XSPEC to estimate the unabsorbed flux in the 0.5--10 keV and 0.1--100 keV energy range. The X-ray luminosities calculated from unabsorbed flux in the 0.1--100 keV energy range are reported in Table \ref{table-para2015}.

The inner disc radius $R_{in}$ can be estimated from \texttt{diskbb} normalization $Norm_{disc}$ using $R_{in}=\alpha \kappa^2 (Norm_{disc}/cos i)^{1/2} D_{10kpc}$ km, where $D_{10kpc}$ is the distance to the source in units of 10 kpc and $i$ is the inclination angle of disc and $\alpha=0.41$ and $\kappa=1.7$ are correction factors \citep{Kubota}. The inner disc radius of $11^{+5}_{-3}$ and $9^{+8}_{-2}$ km is estimated for model A3 and A4, respectively assuming the inclination angle of $80^{\circ}$ \citep[as source is eclipsing binary;][]{Frank}.  The calculated inner disc radius is nearly the order of NS radius or close to NS. 
However, the value of the inner disc radius is uncertain, subjected to the uncertainty in the disc inclination angle $i$. The structure of Comptonized corona was same for both the cases and gave $y-$parameter$\sim 2.7$ and optical depth of $\sim 4.5$.


\subsection{Soft State Emission}
\label{sec-pers2016}


\begin{table*}
\centering
\caption{Spectral parameters of MXB 1658-298 obtained from the different best fit models in high/soft state.}
\resizebox{0.65\linewidth}{!}{
\begin{tabular}{c c c c c c}
\\\hline
Component & parameters & Model B1 & Model B2 & Model B3 &  \\  [0.5ex]
\hline
\textsc{tbabs}	& $N_H (10^{22}~cm^{-2})$	& $0.25 \pm 0.05$ & $<~0.05$ & $0.26 \pm 0.04$ &  \\ [0.5ex]
\hline
\textsc{edge}	& $E_{edge}~(keV)$		& $7.72 \pm 0.05$ & $7.72 \pm 0.05$ & $7.73 \pm 0.05$ & \\ [0.5ex]
		& $\tau$			& $0.162 \pm 0.016$ & $0.153^{+0.019}_{-0.014}$ & $0.153^{+0.017}_{-0.018}$ & \\ [0.5ex]
\hline
\textsc{edge}	& $E_{edge}~(keV)$		& $9.07 \pm 0.07$ & $9.06 \pm 0.07$ & $9.06 \pm 0.07$ & \\ [0.5ex]
		& $\tau$			& $0.129 \pm 0.017$ & $0.110^{+0.008}_{-0.017}$ & $0.119^{+0.017}_{-0.016}$ & \\ [0.5ex]
\hline
\textsc{gauss}	& $E_{line}~(keV)$		& $6.77 \pm 0.03$ & $6.77 \pm 0.03$ & $6.77 \pm 0.03$ & \\ [0.5ex]
		& $width~(keV)$			& $0.09^{+0.05}_{-0.09}$ & $0.098^{+0.05}_{-0.08}$ & $<0.13$ & \\ [0.5ex]
		& Norm ($10^{-4}$)		& $-2.98 \pm 0.5$ & $-3.1 \pm 0.5$ & $-2.87 \pm 0.5$ & \\ [0.5ex]
		& Eq. width (eV)		& $-(52 \pm 6)$ & $-(54^{+8}_{-5})$ & $-(50^{+18}_{-5})$ & \\ [0.5ex]
\hline
\textsc{diskbb} & $kT_{disc}~(keV)$		& & & $0.94^{+0.11}_{-0.12}$ & 		\\ [0.5ex]
		& Norm$_{disc}$			& & & $31^{+14}_{-9}$ &			\\ [0.5ex]
\hline
\textsc{bbodyrad}& $kT_{BB}~(keV)$		& $0.64 \pm 0.04$ & $0.54^{+0.02}_{-0.03}$ &&\\ [0.5ex]
		& Norm$_{BB}$			& $55^{+21}_{-16}$ & $247^{+53}_{-34}$ && \\ [0.5ex]
\hline
\textsc{cutoffpl}& $\Gamma$			& $0.91 \pm 0.08$ &  			\\ [0.5ex]
		& $E_{c}~(keV)$		& $5.3 \pm 0.2$ & 			\\ [0.5ex]
		& Norm				& $0.11 \pm 0.01$ & 			\\ [0.5ex]
\hline
\textsc{nthcomp}& $\Gamma$			& & $2.23^{+0.07}_{-0.06}$ & $2.35 \pm 0.16$ & 			\\ [0.5ex]
		& $kT_{e}~(keV)$		& & $3.80^{+0.15}_{-0.12}$ & $3.97^{+0.33}_{-0.25}$ & 			\\ [0.5ex]
		& $kT_{seed}~(keV)$		& & $0.97^{+0.05}_{-0.09}$ & $1.29^{+0.21}_{-0.24}$ & 			\\ [0.5ex]
		& input\_type			& & 0 &	0 &		\\ [0.5ex]
		& Norm				& & $0.021^{+0.005}_{-0.004}$ & $0.010^{+0.004}_{-0.003}$ & \\ [0.5ex]
\hline
		&$^afrac_{comp}$           & $0.91$ & $0.77$ & $0.56$ & \\ [0.5ex]
 		& $^\dagger$Flux$_{0.5-10~keV}$   & $9.17 \pm 0.22$ & $8.33 \pm 0.03$ & $9.26 \pm 0.04$ & \\ [0.5ex]
 		& $^\dagger$Flux$_{0.1-100~keV}$  & $11.7 \pm 0.3$ & $10.28 \pm 0.04$ & $11.71 \pm 0.05$ & \\ [0.5ex]
 		&$^{\dagger\dagger}L_X$	       & $13.99 \pm 0.36$ & $12.30 \pm 0.05$ & $14.01 \pm 0.06$ & \\ [0.5ex]
\hline
		&$\chi^2/dof$			& 857/840 & 846/839 & 861/839 & \\ [0.5ex]
\hline
\multicolumn{6}{l}{Model B1 : \texttt{TBabs$\times$edge$\times$edge$\times$(bbodyrad+cutoffpl+gaussian)};}\\
\multicolumn{6}{l}{Model B2 : \texttt{TBabs$\times$edge$\times$edge$\times$(bbodyrad+nthcomp[BB]+gaussian)};}\\
\multicolumn{6}{l}{Model B3 : \texttt{TBabs$\times$edge$\times$edge$\times$(diskbb+nthcomp[BB]+gaussian)};}\\
\multicolumn{6}{l}{$^\dagger$Unabsorbed flux in units of $10^{-10}$ \erg.}\\
\multicolumn{6}{l}{$^{\dagger\dagger}$Unabsorbed 0.1--100 keV X-ray luminosity in units of $10^{36}$ erg s$^{-1}$.}\\
\multicolumn{6}{l}{$^a$represents the fraction of Comptonization/cutoffpl component.}\\
\end{tabular}}
\label{table-para2016}
\end{table*}

\begin{figure}
\centering
\includegraphics[width=0.9\columnwidth]{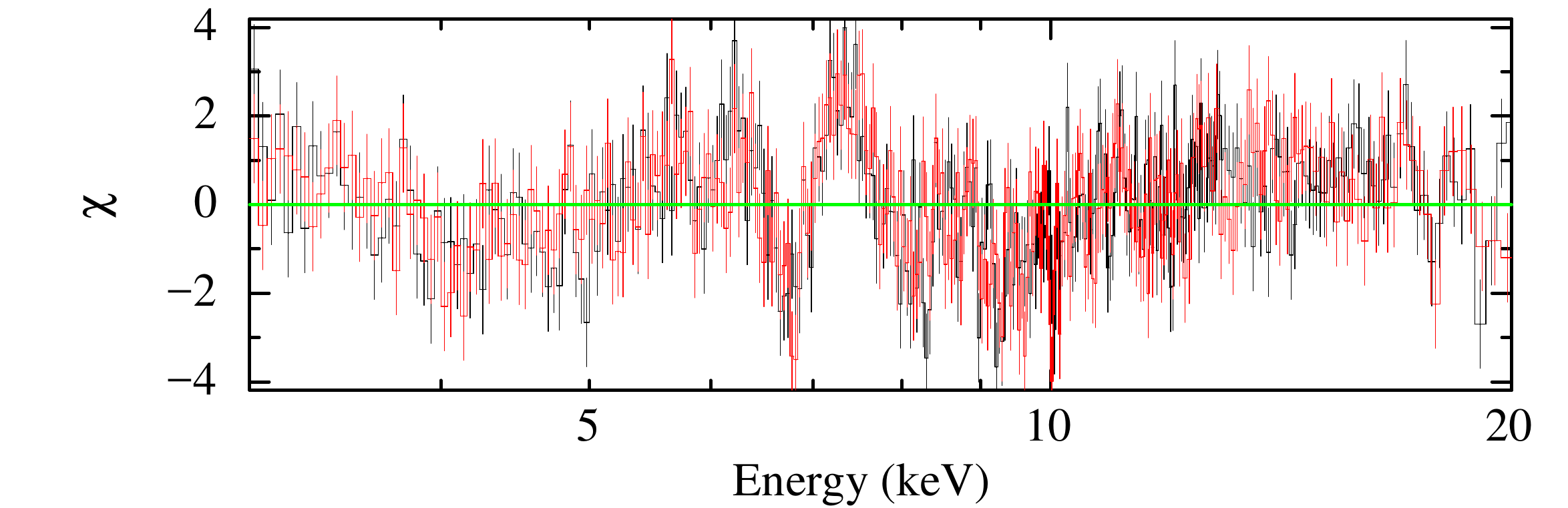}
 \caption{Residuals obtained after modeling persistent spectrum of 2016 observation with absorbed blackbody and cutoff power-law. For simplicity we have shown only \nus\ data in 3--20 keV. These residuals indicate the absorption feature in 6--10 keV (absorption line and two absorption edges) mainly due to Fe K.}
\label{residuals}
\end{figure}

\begin{figure*}
\includegraphics[width=0.69\columnwidth]{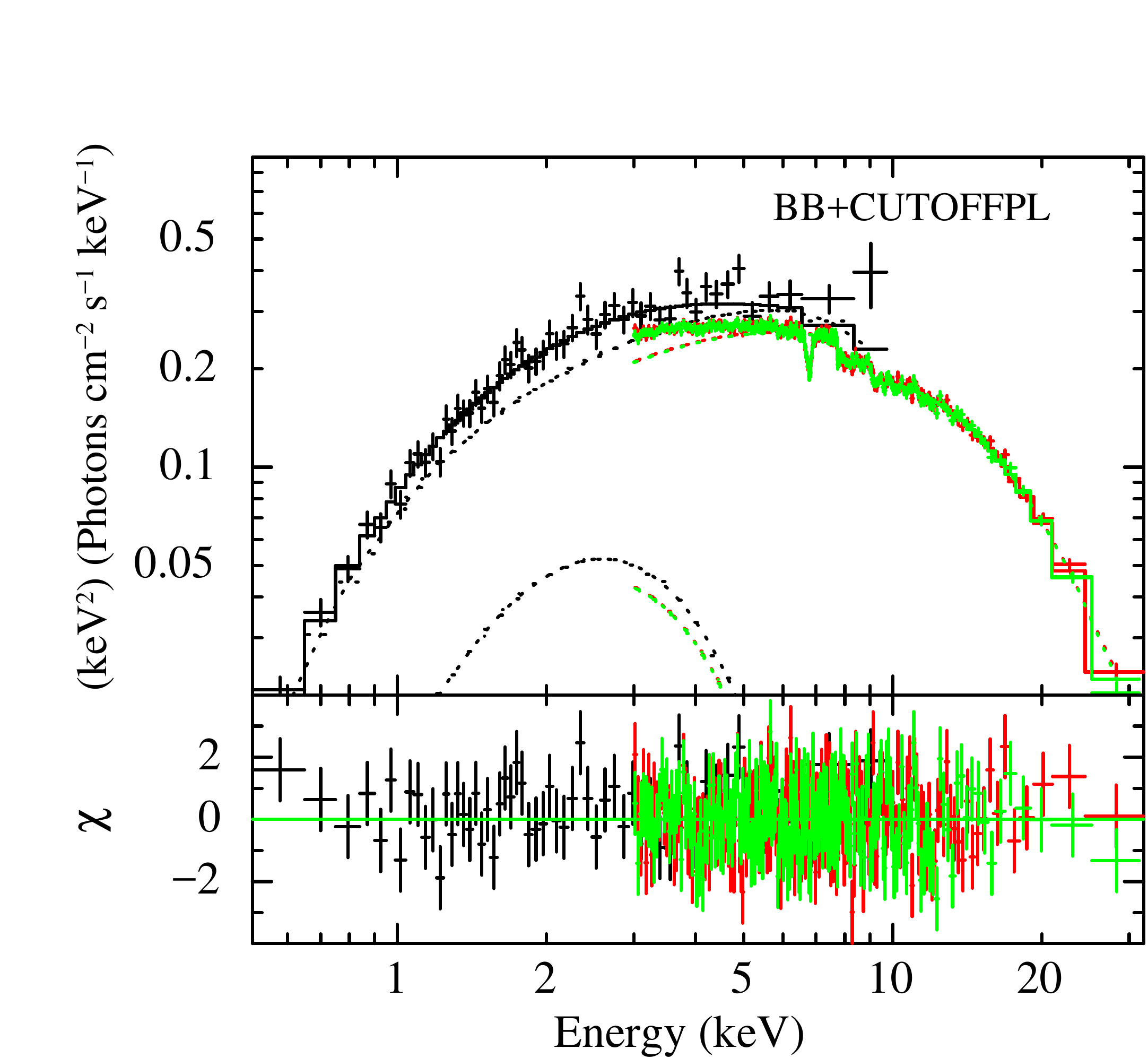}
\includegraphics[width=0.69\columnwidth]{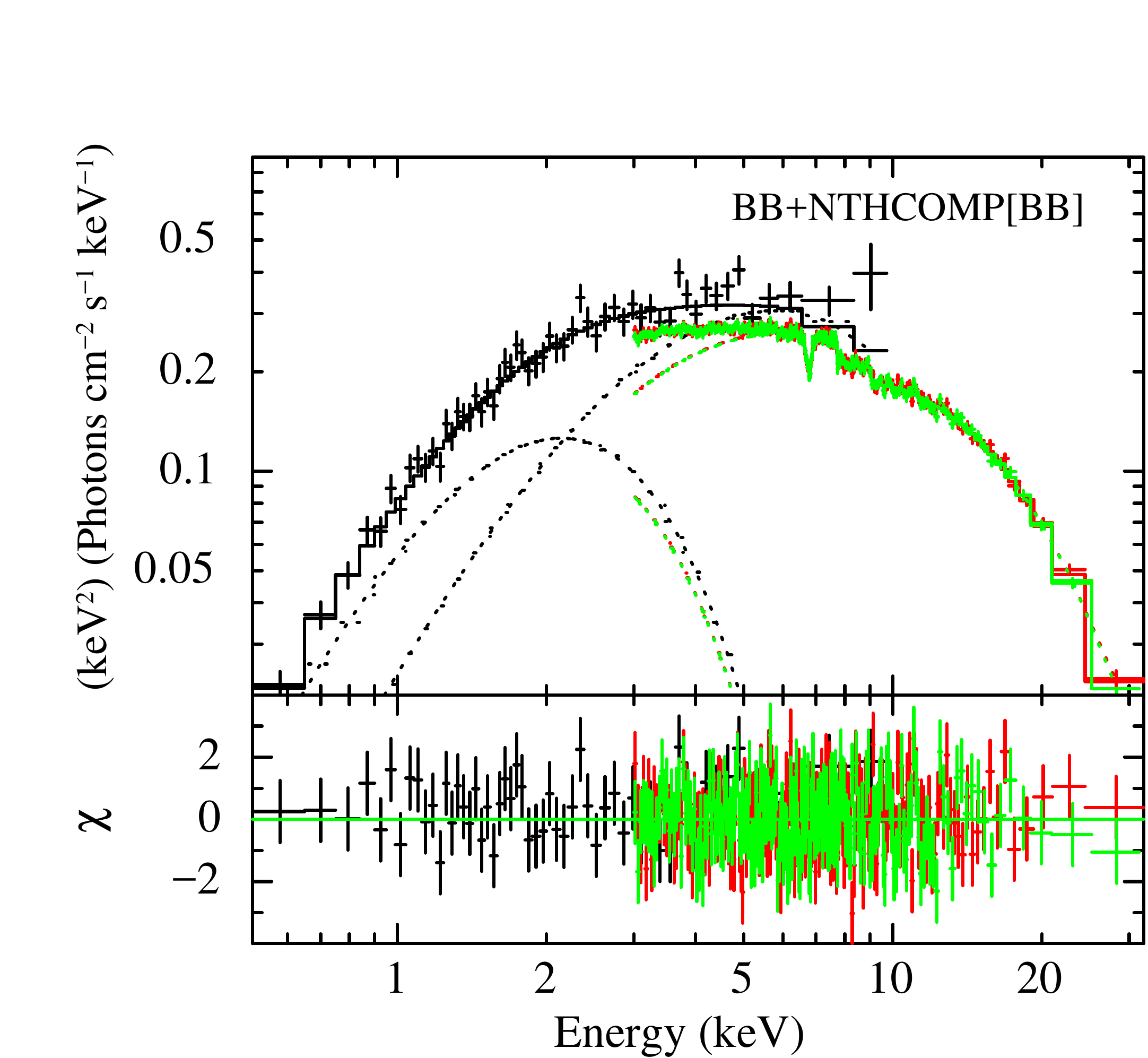}
\includegraphics[width=0.69\columnwidth]{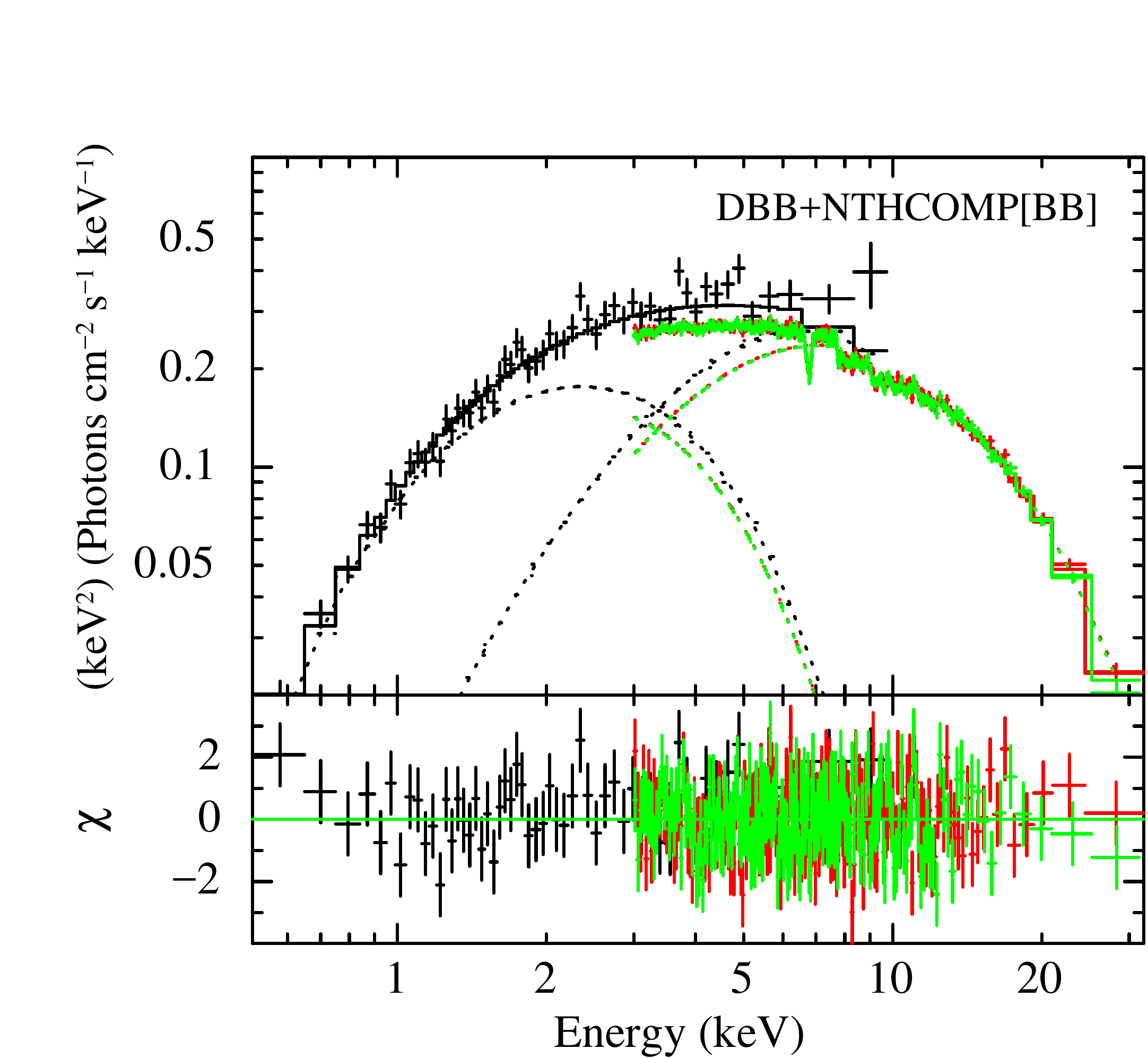}
\caption{ Spectrum of MXB 1658-298 extracted from 2016 observation (soft state) fitted with combinations of thermal and non-thermal components. The model is mentioned on the top-right of each figure. The colour scheme as described in Figure \ref{fig-pe2015}. The figures have been rebinned for representation purpose.}
\label{fig-pe2016}
\end{figure*}


\begin{table}
\centering
\caption{Spectral parameters with single and double absorption detected in the high/soft state, where continuum used was \texttt{TBabs$\times$(bbodyrad+nthcomp[BB])}.}
\resizebox{\columnwidth}{!}{
\begin{tabular}{c c c c}
\\\hline
Component & Parameters & Single Absorption  & Double Absorption \\  [0.5ex]
\hline
\textsc{tbabs}	& $N_H (10^{22}~cm^{-2})$	& $<0.05$ & $<0.044$ \\ [0.5ex]
\hline
\textsc{edge 1}	& $E_{edge}~(keV)$		& $7.72 \pm 0.05$ & $7.70 \pm 0.05$ \\ [0.5ex]
		& $\tau$			& $0.153^{+0.019}_{-0.014}$ & $0.157^{+0.015}_{-0.014}$ \\ [0.5ex]
\hline
\textsc{edge 2}	& $E_{edge}~(keV)$		& $9.06 \pm 0.07$ & $8.83^f$ \\ [0.5ex]
		& $\tau$			& $0.110^{+0.008}_{-0.017}$ & $0.057^{+0.025}_{-0.023}$ \\ [0.5ex]
\hline
\textsc{edge 3}	& $E_{edge}~(keV)$		& - & $9.28^f$ \\ [0.5ex]
		& $\tau$			& - & $0.062^{+0.024}_{-0.021}$ \\ [0.5ex]
\hline
\textsc{gauss 1}	& $E_{line}~(keV)$		& $6.77 \pm 0.03$ & $6.70^f$ \\ [0.5ex]
		& $width~(keV)$			& $0.098^{+0.05}_{-0.08}$ & $0.05~(<0.13)$ \\ [0.5ex]
		& Norm ($10^{-4}$)		& $-3.1 \pm 0.5$ & $-2.45^{+0.38}_{-0.59}$ \\ [0.5ex]
		& Eq. width (eV)		& $-(54^{+8}_{-5})$ & $-(43^{+8}_{-7})$ \\ [0.5ex]
\hline
\textsc{gauss 2}	& $E_{line}~(keV)$		& - & $6.97^f$ \\ [0.5ex]
		& $width~(keV)$			& - & $0.02~(<0.15)$ \\ [0.5ex]
		& Norm ($10^{-4}$)		& - & $-0.93 \pm 0.38$ \\ [0.5ex]
		& Eq. width (eV)		& - & $-(18.6 \pm 7.5)$ \\ [0.5ex]
\hline
\textsc{bbodyrad}	& $kT_{BB}~(keV)$		& $0.54^{+0.02}_{-0.03}$ & $0.55^{+0.02}_{-0.017}$ \\ [0.5ex]
		& Norm$_{BB}$			& $247^{+53}_{-34}$ & $244^{+45}_{-28}$ \\ [0.5ex]
\hline
\textsc{nthcomp}	& $\Gamma$			& $2.23^{+0.07}_{-0.06}$ & $2.23 \pm 0.05$  			\\ [0.5ex]
		& $kT_{e}~(keV)$		& $3.80^{+0.15}_{-0.12}$ & $3.80^{+0.12}_{-0.11}$ 			\\ [0.5ex]
		& $kT_{seed}~(keV)$		& $0.97^{+0.05}_{-0.09}$ & $0.99^{+0.10}_{-0.05}$  			\\ [0.5ex]
		& input\_type			& 0 & 0 \\ [0.5ex]
		& Norm				& $0.021^{+0.005}_{-0.004}$ & $0.020^{+0.003}_{-0.004}$  \\ [0.5ex]
\hline
		&$\chi^2/dof$			& 846/839 & 855/838  \\ [0.5ex]
\hline
\multicolumn{4}{l}{$^f$ Fixed.} \\

\end{tabular}}
\label{table-abs}
\end{table}

During the observation of 2016, \mxb\ was in the high/soft state and significantly detected above background upto 35 keV only. We have modeled the persistent spectra of 2016 observation with absorbed cutoff power-law with a thermal blackbody component, \texttt{TBabs*(bbodyrad+cutoffpl)}. The fit obtained was unacceptable with $\chi^2/dof=1290/847$. This model described the 0.5--35 keV continuum well except between 6--10 keV. The structure of residuals showed the presence of an absorption line near 6.7 keV and two absorption edges at 7.7 keV and 9 keV (see, Figure \ref{residuals}). So we added a Gaussian absorption line model (\texttt{gauss}) and two absorption edge models (\texttt{edge}). This resulted in an acceptable fit with $\chi^2/dof=857/840$ (fit improvement by adding an edge at 9 keV: $-\Delta \chi^2=121$ for 2 additional parameters, another edge at 7.7 keV: $-\Delta \chi^2=143.7$ for 2 additional parameters and Gaussian absorption line at 6.7 keV: $-\Delta \chi^2=167.7$ for 3 additional parameters). The best fit parameters of this model (B1) are given in Table \ref{table-para2016}. 
With this model, the photon index obtained was hard ($\sim 0.91$). Similar hard photon index during soft state was found by \citet{Bloser2000, Bloser2000b} with cutoff power-law model. If \texttt{cutoffpl} is replaced with \texttt{powerlaw}, the spectrum upto 10 keV will give photon index of $\sim 1.8$, consistent with the value measured in the soft state of NS LMXBs \citep{Boirin2005, Trigo, Raman2018}.

In the next approach, we replaced the cutoff power-law component with Comptonization component \texttt{nthcomp}. The model \texttt{TBabs*(bbodyrad+nthcomp[BB or diskbb])} also showed the presence of same spectral feature of absorption in 6--10 keV. So, we added the same Gaussian absorption line and two absorption edge models. 
The model with \texttt{nthcomp[diskbb]} resulted in unphysical values of emission radius of disc photons $\sim 2-3$ km and $kT_{BB}<kT_{disc}$ \citep{Lin07}. When input seed photon type changed to blackbody (\texttt{nthcomp[BB]}; model B2), an acceptable fit was obtained (Table \ref{table-para2016}). 

We replaced the \texttt{bbodyrad} with \texttt{diskbb} in model B2. This gave a marginally poor fit but it describes the continuum well. This model (B3) represents the blackbody emission from NS surface/boundary layer with $kT_{BB}=1.29^{+0.21}_{-0.24}$ keV and emission radius of $R_0\sim 3-4$ km is completely Comptonized and emission from accretion disc with $kT_{disc}=0.94^{+0.11}_{-0.12}$ keV and inner disc radius of $R_{in}=15.8^{+3.7}_{-2.3}$ km is directly visible. 
We have also found that adding a third thermal component, \texttt{diskbb} in model B2 and \texttt{bbodyrad} in B3 did not improve the fit. So, the continuum during high/soft state can be described with combination of blackbody or disc blackbody and thermally Comptonization component where input seed photons are provided by blackbody (NS surface/boundary layer). 
Figure \ref{fig-pe2016} shows the best fit spectrum of \mxb\ in the soft state from above described  best fit models B1, B2 and B3. The model \texttt{TBabs $ \times $ (bbodyrad+nthcomp[BB])} i.e., model B2 reproduce the data best as it improve the residuals in the soft band.
The best fit parameters and the X-ray luminosities calculated from unabsorbed flux in the 0.1--100 keV energy range are reported in Table \ref{table-para2016}. The estimated X-ray luminosity during 2016 was a factor of $\sim 4$ higher than 2015.

\begin{figure}
\centering
\includegraphics[width=0.8\columnwidth]{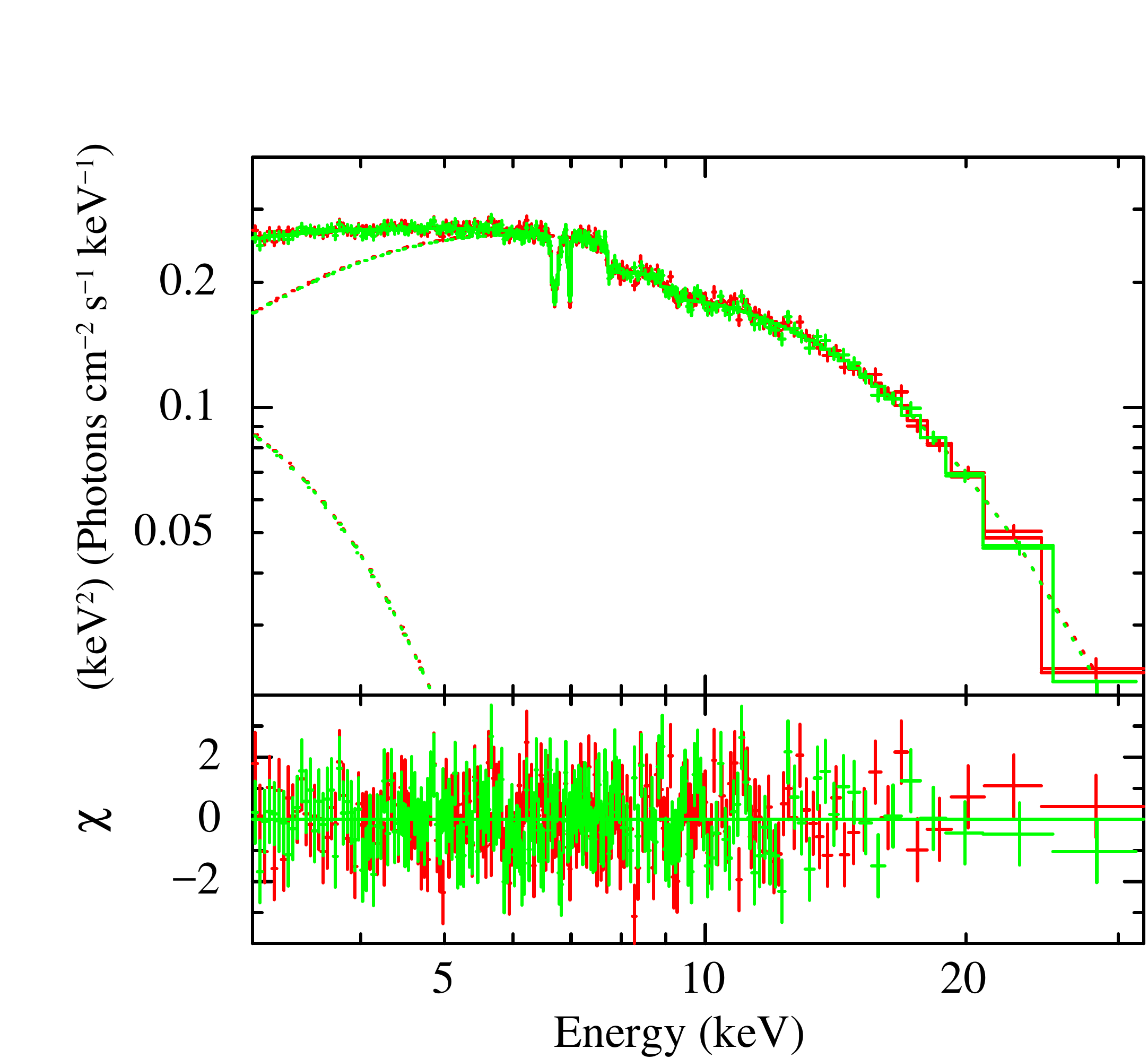}
 \caption{Spectrum of \mxb\ during high/soft state when modeled with double absorption. For representation purpose, figure has been rebinned and only \nus\ data is shown.}
\label{unfolded}
\end{figure}

\subsubsection{Absorption features}

The absorption line observed during soft state at $6.77 \pm 0.03$ keV showed line width of $ \sim 0.1$ keV and was broader than the results of the previous observations, when single absorption lines from Fe \textsc{xxv} and Fe \textsc{xxvi} were detected \citep{Sidoli}. This line can be due to blend of Fe \textsc{xxv} K$\alpha$ (6.70 keV) and Fe \textsc{xxvi} K$\alpha$ (6.97 keV) absorption lines. Accordingly, we applied two Gaussian absorption lines at 6.70 and 6.97 keV by freezing the line energy parameter instead of single Gaussian absorption line in the model B2. Both lines obtained were unresolved, we found the upper limits on the line widths of <0.12 keV and <0.14 keV for Fe \textsc{xxv} and Fe \textsc{xxvi}, respectively. The equivalent width (EW) of Fe \textsc{xxv} and \textsc{xxvi} K$\alpha$ lines obtained were $-42 \pm 8$ eV and $-18 \pm 7$ eV, respectively. 
The two absorption edges were observed at $7.72 \pm 0.05$ keV and $9.06 \pm 0.07$ keV with optical depth of $\sim 0.15$ and $0.11$, respectively. The absorption edge at 9 keV is due to highly ionized Fe \textsc{xxv}/\textsc{xxvi} K edge. 
Similar, absorption feature as observed in \mxb\ were also observed in Galactic jet source GRO J1655--40 \citep{Yamaoka}. 
Since absorption line features of Fe K ions are detected, the accompanying absorption edge structures from the Fe ions in the same ionization states would also appear in the spectra. Hence following \citet{Yamaoka} and \citet{Sidoli2002}, we applied two absorption edges fixed at 8.83 keV (Fe \textsc{xxv}) and 9.28 keV (Fe \textsc{xxvi}) instead of the single 9 keV edge detected above. The best-fit parameters for single and double absorptions are shown in Table \ref{table-abs}. The obtained optical depth for 8.83 keV and 9.28 keV edges were nearly same $\sim 0.06$, shows both edges arises from nearly same region. Figure \ref{unfolded} shows the spectra with double absorption lines from Fe \textsc{xxv} and \textsc{xxvi} K$\alpha$ and their respective absorption edges.

The absorption lines due to Fe \textsc{xxv} K$\beta$ and Fe \textsc{xxvi} K$\beta$ at 7.88 keV and 8.26 keV, respectively had been observed in NS LMXB AX J1745.6--2901 \citep{Hyodo2009,PontiAX-2015} and GX 13+1 \citep{Sidoli2002}. The observed absorption edge at 7.7 keV can be due to Fe \textsc{xxv/xxvi} K$\beta$ absorption lines with contribution from highly ionized Ni K$\alpha$ absorptions lines \citep{Yamaoka} or(/and) Fe K emission as absorption edge can mimic the broad Fe K emission line \citep{DAi, Egron, Mondal2016}


\subsection{Photo-ionized Absorption Model}

\begin{figure}
\centering
\includegraphics[width=0.8\columnwidth]{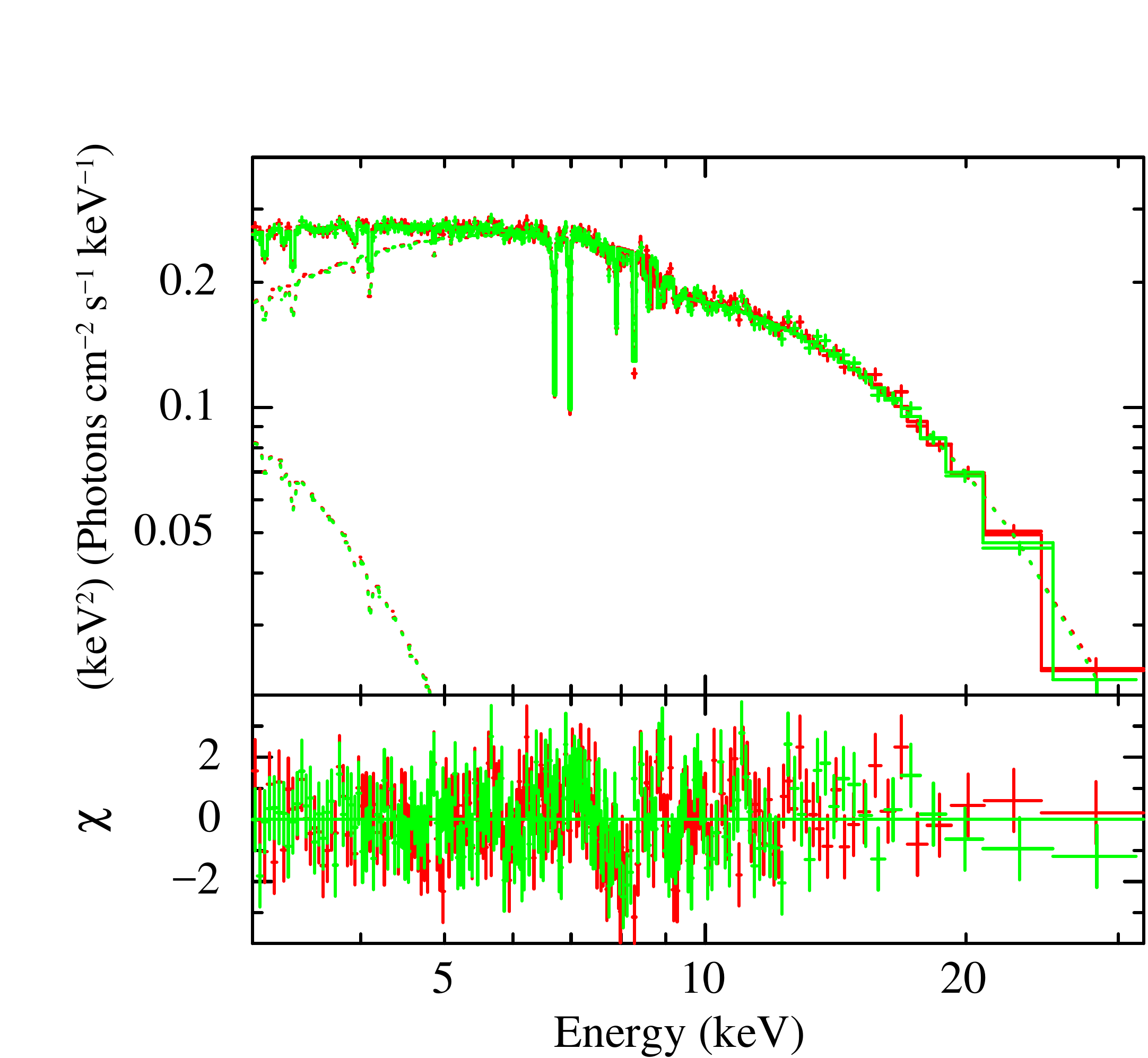}
 \caption{Spectrum when modeled with \texttt{TBabs*zxipcf*(bbodyrad+nthcomp[BB])}. For representation purpose, figure has been rebinned and only \nus\ data is shown.}
\label{zxipcf}
\end{figure}


We have also modeled the observed absorption features in the soft state with a photo-ionized absorption component \citep[\texttt{zxipcf};][]{Reeves} and assumed to be totally covering the X-ray source (covering fraction was fixed to 1). We modeled the broadband spectrum with model \texttt{Tbabs*zxipcf*(bbodyrad+nthcomp[BB])}. We note that \texttt{zxipcf} model was able to reproduce Fe K absorption lines between 6--7 keV and absorption edge at 9 keV (as shown in Figure \ref{zxipcf}). However residuals at 7--8 keV were still present. The fit showed positive residuals around 7 keV and negative residuals at 8 keV (see lower panel of Figure \ref{zxipcf}). The fit obtained was not good, $\chi^2/dof=927/844$. 
We first modeled these 7--8 keV features with an absorption edge. The fit improved to $\chi^2/dof=876.5/842$ with column density of $N_H \sim 10^{24}$ cm$^{-2}$ and ionization parameter of log$(\xi) \sim 4.2$ of ionized absorber. The absorption edge was observed at $7.53 \pm 0.09$ keV with optical depth of $0.07 \pm 0.01$. In the second approach, we modeled this feature with emission line using Gaussian model and fit improved to $\chi^2/dof=878/841$. The line energy observed was $6.78^{+0.06}_{-0.09}$ keV with line width of $0.38^{+0.11}_{-0.04}$ keV. 
The best fitting column density of ionized layer was $N_H = 6.7^{+1.2}_{-2.4} \times 10^{23}$ cm$^{-2}$, having an ionization parameter log$(\xi) = 4.06^{+0.09}_{-0.14}$. The EW of Fe K emission line obtained was $61^{+27}_{-20}$ eV. The edge at 7.5 keV can mimic the Fe K emission feature which was also found in 4U 1728--34 \citep{DAi,Egron} and 4U 1820--30 \citep{Mondal2016}. Broad Fe K emission line was also observed during previous outburst with EW of $41^{+17}_{-9}$ eV in \emph{BeppoSAX} spectra \citep{Oosterbroek} and $160^{+60}_{-40}$ eV in \xm\ spectra \citep{Sidoli}. 

\subsubsection{Absorption in Hard State}

Since no absorption feature were detected during persistent spectrum of 2015, therefore we added the Gaussian absorption line in model A1 to estimate upper limits. 
We found an absorption line at $6.63^{+0.13}_{-0.12}$ keV (line width fixed to 10 eV) with EW of $-15 \pm 6$ eV with marginal fit improvement of $\Delta \chi^2 = 6$ for 2 additional parameters with F-test probability of 0.07. The addition of absorption line did not improve the fit significantly. So, this absorption feature is not much significant in this 2015 observation. We also estimated the upper limit on the EW of 6.4 keV Fe emission line of $16.5$ eV (line width fixed to 0.3 keV) and on 6.97 keV Fe \textsc{xxvi} absorption line of $-9$ eV (line width fixed to 10 eV). We have also modeled the persistent spectrum of 2015 with photo-ionization absorption model \texttt{zxipcf}. We used the resultant model \texttt{Tbabs $\times$ zxipcf $\times$ (boodyrad + nthcomp[diskbb])}. The covering fraction and redshift was fixed to 1 and zero, respectively. This improved the fit with $\Delta \chi^2 = $4 for 2 additional parameters with F-test probability of 0.16, which is not significant. So during observation of 2015, no absorption features were detected significantly as observed in the observation of 2016.


\section{Discussion and Conclusion}

This work shows the comparison of the broadband spectrum of \mxb\ in its soft and hard states. This kind of analysis has been done for the first time in this source. During its recent outburst, it was in low/hard state during the 2015 observation and in high/soft states during the 2016 observation. 
Many of the NS LMXBs have shown the transitions between the spectral states \citep{Munoz}. Transient atoll source Aql X--1 has been observed in both hard and soft states during its 2007 outburst \citep{Raichur2011,Sakurai2012}. 
During the outburst of 2011, Aql X--1 is reported to show a spectral transition from hard to soft state \citep{Ono2017}. 
The transient LMXB AX J1745.6--2901 is also known to show both the spectral states during an outburst \citep{PontiAX}.

During low/hard state, the combined broadband spectrum from \nus\ and \sw\ observations can be well described with combination of thermal component (blackbody) and Comptonized disc component. However, combination of two thermal components (blackbody and disc blackbody) and Comptonized component can not be ruled out (see, Table \ref{table-para2015}). 
During 2016 observation when source was in high/soft state, its broadband spectrum can be described with combination of blackbody or disc blackbody and a Comptonization component (see, Table \ref{table-para2016}). In the soft state, our study supports the NS surface (or boundary layer) as the dominant source for the Comptonization seed photons yielding the observed weak hard emission, while in the hard state inner accretion disc is favoured but with three-component model solutions, either the disc or the NS surface/boundary layer are equally favoured.
The spectral shape during both observations were different and were according to the spectral state of the source. The electron temperature of Comptonizing plasma decreased to $\sim 4$ keV in high/soft state from $\sim 19$ keV in low/hard state. During low/hard state, Comptonizing region had optical depth of $\sim 4.1$ and during high/soft state, plasma became optically thick with optical depth of $\sim$ 7. This type of variations in electron temperature and plasma optical depth have been observed during the hard and soft spectral state of several atoll LMXBs \citep[e.g.,][]{Done, Raichur2011, Sakurai2012, Ono2017}. 

Like \mxb, the broadband spectrum of Aql X--1, a typical non-dipping LMXB and 4U 1915--05, a typical dipping LMXB were also modeled with \texttt{diskbb+nthcomp[BB]} model in the soft state \citep{Sakurai2012,Zhang2014}.
They showed blackbody emission of $R_{BB} \sim 3-4$ km, which implies that the blackbody emission arises from an equatorial belt-like region on the NS surface. They also commonly indicate that the inner edge of the accretion disc is close to the NS surface ($R_{in} \sim 10-20$ km). The measured coronal temperature $kT_e$ of \mxb\ $\sim 4$ keV, is somewhat higher than that of Aql X--1 ($2-3$ keV), but is significantly lower than 4U 1915--05 ($\sim 9$ keV) \citep{Sakurai2012, Zhang2014}. The dipping LMXBs in soft state (banana state) have systematically harder spectra than the normal LMXBs \citep{Gladstone2007}. 
The strength of Comptonization is normally evaluated by the $y-$parameter. 
Following \citet{Zhang2014}, for LMXBs in the soft state, the difference between $kT_e$ and $kT_{seed}$ becomes small and they employed a new definition of the $y-$parameter as $4 (\tau + \tau^2/3) (kT_e - kT_{seed})/m_e c^2$.  
The newly defined $y-$parameter of \mxb\ is $\sim 0.5$ similar to 4U 1915--05 in the soft state \citep{Zhang2014}. This value appears at the upper limit of the recalculated $y-$parameters of normal LMXBs in the soft state by \citet{Zhang2014}. 
Thus similar to 4U 1915--05, \mxb\ is inferred to show stronger Comptonization effects than low and medium inclination LMXBs. 
We have obtained $y-$parameter $\sim 2.5$ in the hard state (as defined in section \ref{hardstate}), a value higher than EXO 0748--676 ($y \sim 1.4$) and other low and medium inclined LMXBs \citep[$y < 1.4$;][]{Zhang2016}. 
The large difference in the value of $y-$parameter between \mxb\ and EXO 0748--676 (both dipping LMXBs) can be due to the spectral modeling. EXO 0748--676 was modeled with \texttt{diskbb+nthcomp[BB]} \citep{Zhang2016} while \mxb\ was modeled with \texttt{bbodyrad+nthcomp[diskbb]}. Our model assume the the disc is fully covered by Comptonizing corona and the high value of $y-$parameter suggests that corona is flattened over the disc.
So, our study suggests that the Comptonizing corona has an oblate shape and the seed photons have to pass through the corona with a longer path resulting in systematically stronger Comptonization \citep{Zhang2014, Zhang2016}. 

We have observed absorption lines due to highly ionized Fe K in \mxb\ during high/soft state. The Fe \textsc{xxv} K$\alpha$ (6.70 keV) and Fe \textsc{xxvi} K$\alpha$ (6.97 keV) absorption lines were observed with EW of  $-43^{+8}_{-7}$ eV and $-18.6 \pm 7.5$ eV, respectively with their corresponding absorption edges with optical depth of $\sim 0.06$. For the first time absorption edges due to highly ionized Fe is observed in \mxb. 
The intense Fe K absorption features during soft states have been found in both NS and BH systems \citep[e.g.,][]{Neilsen2009, Ponti12, Ponti14, Ponti2016, PontiAX, Bozzo, Degenaar2014, Degenaar2016}. The Fe K absorber is related to the presence of accretion disc absorber. In nearly 70\% of the known NS LMXBs, absorbing material is known to be bound to the accretion disc atmosphere. There is no shift in their absorption line energies. The rest of systems show blueshift in their absorption line energies \citep{Trigo2013}. For example, \mxb\ (previous outburst), 4U 1323--62 and EXO 0748--676 did not show significant energy shifts, indicating that the absorption might occur in the disc atmosphere and does not necessarily require an outflow \citep{Sidoli,Trigo,Boirin2005,Ponti14}. 
So, the observed absorption lines and edges in \mxb\ are probably due to accretion disc atmosphere (or wind) as the source is viewed close to edge-on. Since, the observed K-absorption lines and K-absorption edges due to Fe ions were absent or not significantly detected during low/hard state. So, this suggests a connection between Fe K absorber and spectral state of the source. The connection between spectral state and presence of accretion disc wind or atmosphere have been found in both BH systems \citep{Neilsen2009,Ponti12} and high inclination NS LMXBs: EXO 0748--676 \citep{Ponti14} and AX J1745.6--2901 \citep{PontiAX-2015, PontiAX}. Our study presents an overall picture of high inclination LMXBs (BH and NS) and their spectral states/disc absorber connection. 

\section*{Acknowledgements}

This research has made use of data obtained from the High Energy Astrophysics Science Archive Research Center (HEASARC), provided by NASA's Goddard Space Flight Center. 
This research has made use of the NuSTAR Data Analysis Software (NUSTARDAS) jointly developed by the ASI Science Data Center (ASDC, Italy) and the California Institute of Technology (USA). RS and AJ acknowledge the financial support from the University Grants Commission (UGC), India, under the SRF and MANF-SRF schemes, respectively. 




\bibliographystyle{mnras}

\begin{thebibliography}{}
\makeatletter
\relax
\def\mn@urlcharsother{\let\do\@makeother \do\$\do\&\do\#\do\^\do\_\do\%\do\~}
\def\mn@doi{\begingroup\mn@urlcharsother \@ifnextchar [ {\mn@doi@}
  {\mn@doi@[]}}
\def\mn@doi@[#1]#2{\def\@tempa{#1}\ifx\@tempa\@empty \href
  {http://dx.doi.org/#2} {doi:#2}\else \href {http://dx.doi.org/#2} {#1}\fi
  \endgroup}
\def\mn@eprint#1#2{\mn@eprint@#1:#2::\@nil}
\def\mn@eprint@arXiv#1{\href {http://arxiv.org/abs/#1} {{\tt arXiv:#1}}}
\def\mn@eprint@dblp#1{\href {http://dblp.uni-trier.de/rec/bibtex/#1.xml}
  {dblp:#1}}
\def\mn@eprint@#1:#2:#3:#4\@nil{\def\@tempa {#1}\def\@tempb {#2}\def\@tempc
  {#3}\ifx \@tempc \@empty \let \@tempc \@tempb \let \@tempb \@tempa \fi \ifx
  \@tempb \@empty \def\@tempb {arXiv}\fi \@ifundefined
  {mn@eprint@\@tempb}{\@tempb:\@tempc}{\expandafter \expandafter \csname
  mn@eprint@\@tempb\endcsname \expandafter{\@tempc}}}

\bibitem[\protect\citeauthoryear{{Armas Padilla}, {Ueda}, {Hori}, {Shidatsu}
  \& {Mu{\~n}oz-Darias}}{{Armas Padilla} et~al.}{2017}]{Armas}
{Armas Padilla} M.,  {Ueda} Y.,  {Hori} T.,  {Shidatsu} M.,
  {Mu{\~n}oz-Darias} T.,  2017, \mn@doi [\mnras] {10.1093/mnras/stx020}, \href
  {http://adsabs.harvard.edu/abs/2017MNRAS.467..290A} {467, 290}

\bibitem[\protect\citeauthoryear{{Arnaud}}{{Arnaud}}{1996}]{Arnaud}
{Arnaud} K.~A.,  1996, in {Jacoby} G.~H.,  {Barnes} J.,  eds,  Astronomical
  Society of the Pacific Conference Series Vol. 101, Astronomical Data Analysis
  Software and Systems V. p.~17

\bibitem[\protect\citeauthoryear{{Bahramian}, {Heinke}  \&
  {Wijnands}}{{Bahramian} et~al.}{2015}]{Bahramian}
{Bahramian} A.,  {Heinke} C.~O.,   {Wijnands} R.,  2015, The Astronomer's
  Telegram, \href {http://adsabs.harvard.edu/abs/2015ATel.7957....1B} {7957}

\bibitem[\protect\citeauthoryear{{Bloser}, {Grindlay}, {Barret}  \&
  {Boirin}}{{Bloser} et~al.}{2000a}]{Bloser2000}
{Bloser} P.~F.,  {Grindlay} J.~E.,  {Barret} D.,   {Boirin} L.,  2000a, \mn@doi
  [\apj] {10.1086/317035}, \href
  {http://adsabs.harvard.edu/abs/2000ApJ...542..989B} {542, 989}

\bibitem[\protect\citeauthoryear{{Bloser}, {Grindlay}, {Kaaret}, {Zhang},
  {Smale}  \& {Barret}}{{Bloser} et~al.}{2000b}]{Bloser2000b}
{Bloser} P.~F.,  {Grindlay} J.~E.,  {Kaaret} P.,  {Zhang} W.,  {Smale} A.~P.,
  {Barret} D.,  2000b, \mn@doi [\apj] {10.1086/317019}, \href
  {http://adsabs.harvard.edu/abs/2000ApJ...542.1000B} {542, 1000}

\bibitem[\protect\citeauthoryear{{Boirin}, {Parmar}, {Barret}  \&
  {Paltani}}{{Boirin} et~al.}{2004}]{Boirin2004}
{Boirin} L.,  {Parmar} A.~N.,  {Barret} D.,   {Paltani} S.,  2004, \mn@doi
  [Nuclear Physics B Proceedings Supplements]
  {10.1016/j.nuclphysbps.2004.04.085}, \href
  {http://adsabs.harvard.edu/abs/2004NuPhS.132..506B} {132, 506}

\bibitem[\protect\citeauthoryear{{Boirin}, {M{\'e}ndez}, {D{\'{\i}}az Trigo},
  {Parmar}  \& {Kaastra}}{{Boirin} et~al.}{2005}]{Boirin2005}
{Boirin} L.,  {M{\'e}ndez} M.,  {D{\'{\i}}az Trigo} M.,  {Parmar} A.~N.,
  {Kaastra} J.~S.,  2005, \mn@doi [\aap] {10.1051/0004-6361:20041940}, \href
  {http://adsabs.harvard.edu/abs/2005A%26A...436..195B} {436, 195}

\bibitem[\protect\citeauthoryear{{Bozzo} et~al.,}{{Bozzo} et~al.}{2016}]{Bozzo}
{Bozzo} E.,  et~al., 2016, \mn@doi [\aap] {10.1051/0004-6361/201527501}, \href
  {http://adsabs.harvard.edu/abs/2016A%26A...589A..42B} {589, A42}

\bibitem[\protect\citeauthoryear{{Burrows} et~al.,}{{Burrows}
  et~al.}{2005}]{Burrows}
{Burrows} D.~N.,  et~al., 2005, \mn@doi [\ssr] {10.1007/s11214-005-5097-2},
  \href {http://adsabs.harvard.edu/abs/2005SSRv..120..165B} {120, 165}

\bibitem[\protect\citeauthoryear{{Cominsky} \& {Wood}}{{Cominsky} \&
  {Wood}}{1984}]{CominskynWood}
{Cominsky} L.~R.,  {Wood} K.~S.,  1984, \mn@doi [\apj] {10.1086/162361}, \href
  {http://adsabs.harvard.edu/abs/1984ApJ...283..765C} {283, 765}

\bibitem[\protect\citeauthoryear{{D'A{\'{\i}}} et~al.,}{{D'A{\'{\i}}}
  et~al.}{2006}]{DAi}
{D'A{\'{\i}}} A.,  et~al., 2006, \mn@doi [\aap] {10.1051/0004-6361:20053228},
  \href {http://adsabs.harvard.edu/abs/2006A%26A...448..817D} {448, 817}

\bibitem[\protect\citeauthoryear{{Degenaar}, {Miller}, {Harrison}, {Kennea},
  {Kouveliotou}  \& {Younes}}{{Degenaar} et~al.}{2014}]{Degenaar2014}
{Degenaar} N.,  {Miller} J.~M.,  {Harrison} F.~A.,  {Kennea} J.~A.,
  {Kouveliotou} C.,   {Younes} G.,  2014, \mn@doi [\apjl]
  {10.1088/2041-8205/796/1/L9}, \href
  {http://adsabs.harvard.edu/abs/2014ApJ...796L...9D} {796, L9}

\bibitem[\protect\citeauthoryear{{Degenaar} et~al.,}{{Degenaar}
  et~al.}{2016}]{Degenaar2016}
{Degenaar} N.,  et~al., 2016, \mn@doi [\mnras] {10.1093/mnras/stw1593}, \href
  {http://adsabs.harvard.edu/abs/2016MNRAS.461.4049D} {461, 4049}

\bibitem[\protect\citeauthoryear{{D{\'{\i}}az Trigo} \& {Boirin}}{{D{\'{\i}}az
  Trigo} \& {Boirin}}{2013}]{Trigo2013}
{D{\'{\i}}az Trigo} M.,  {Boirin} L.,  2013, Acta Polytechnica, \href
  {http://adsabs.harvard.edu/abs/2013AcPol..53..659D} {53, 659}

\bibitem[\protect\citeauthoryear{{D{\'{\i}}az Trigo} \& {Boirin}}{{D{\'{\i}}az
  Trigo} \& {Boirin}}{2016}]{Trigo2016}
{D{\'{\i}}az Trigo} M.,  {Boirin} L.,  2016, \mn@doi [Astronomische
  Nachrichten] {10.1002/asna.201612315}, \href
  {http://adsabs.harvard.edu/abs/2016AN....337..368D} {337, 368}

\bibitem[\protect\citeauthoryear{{D{\'{\i}}az Trigo}, {Parmar}, {Boirin},
  {M{\'e}ndez}  \& {Kaastra}}{{D{\'{\i}}az Trigo} et~al.}{2006}]{Trigo}
{D{\'{\i}}az Trigo} M.,  {Parmar} A.~N.,  {Boirin} L.,  {M{\'e}ndez} M.,
  {Kaastra} J.~S.,  2006, \mn@doi [\aap] {10.1051/0004-6361:20053586}, \href
  {http://adsabs.harvard.edu/abs/2006A%26A...445..179D} {445, 179}

\bibitem[\protect\citeauthoryear{{Done}, {Gierli{\'n}ski}  \& {Kubota}}{{Done}
  et~al.}{2007}]{Done}
{Done} C.,  {Gierli{\'n}ski} M.,   {Kubota} A.,  2007, \mn@doi [\aapr]
  {10.1007/s00159-007-0006-1}, \href
  {http://adsabs.harvard.edu/abs/2007A%26ARv..15....1D} {15, 1}

\bibitem[\protect\citeauthoryear{{Egron} et~al.,}{{Egron} et~al.}{2011}]{Egron}
{Egron} E.,  et~al., 2011, \mn@doi [\aap] {10.1051/0004-6361/201016093}, \href
  {http://adsabs.harvard.edu/abs/2011A%26A...530A..99E} {530, A99}

\bibitem[\protect\citeauthoryear{{Frank}, {King}  \& {Lasota}}{{Frank}
  et~al.}{1987}]{Frank}
{Frank} J.,  {King} A.~R.,   {Lasota} J.-P.,  1987, \aap, \href
  {http://adsabs.harvard.edu/abs/1987A%26A...178..137F} {178, 137}

\bibitem[\protect\citeauthoryear{{Gehrels} et~al.,}{{Gehrels}
  et~al.}{2004}]{Gehrels}
{Gehrels} N.,  et~al., 2004, \mn@doi [\apj] {10.1086/422091}, \href
  {http://adsabs.harvard.edu/abs/2004ApJ...611.1005G} {611, 1005}

\bibitem[\protect\citeauthoryear{{Gladstone}, {Done}  \&
  {Gierli{\'n}ski}}{{Gladstone} et~al.}{2007}]{Gladstone2007}
{Gladstone} J.,  {Done} C.,   {Gierli{\'n}ski} M.,  2007, \mn@doi [\mnras]
  {10.1111/j.1365-2966.2007.11675.x}, \href
  {http://adsabs.harvard.edu/abs/2007MNRAS.378...13G} {378, 13}

\bibitem[\protect\citeauthoryear{{Harrison} et~al.,}{{Harrison}
  et~al.}{2013}]{Harrison}
{Harrison} F.~A.,  et~al., 2013, \mn@doi [\apj] {10.1088/0004-637X/770/2/103},
  \href {http://adsabs.harvard.edu/abs/2013ApJ...770..103H} {770, 103}

\bibitem[\protect\citeauthoryear{{Hasinger} \& {van der Klis}}{{Hasinger} \&
  {van der Klis}}{1989}]{Hasinger}
{Hasinger} G.,  {van der Klis} M.,  1989, \aap, \href
  {http://adsabs.harvard.edu/abs/1989A%26A...225...79H} {225, 79}

\bibitem[\protect\citeauthoryear{{Hyodo}, {Ueda}, {Yuasa}, {Maeda}, {Makishima}
   \& {Koyama}}{{Hyodo} et~al.}{2009}]{Hyodo2009}
{Hyodo} Y.,  {Ueda} Y.,  {Yuasa} T.,  {Maeda} Y.,  {Makishima} K.,   {Koyama}
  K.,  2009, \mn@doi [\pasj] {10.1093/pasj/61.sp1.S99}, \href
  {http://adsabs.harvard.edu/abs/2009PASJ...61S..99H} {61, S99}

\bibitem[\protect\citeauthoryear{{Iaria} et~al.,}{{Iaria}
  et~al.}{2016}]{Iaria2016}
{Iaria} R.,  et~al., 2016, \mn@doi [\aap] {10.1051/0004-6361/201628210}, \href
  {http://adsabs.harvard.edu/abs/2016A%26A...596A..21I} {596, A21}

\bibitem[\protect\citeauthoryear{{Jain}, {Paul}, {Sharma}, {Jaleel}  \&
  {Dutta}}{{Jain} et~al.}{2017}]{Jain2017}
{Jain} C.,  {Paul} B.,  {Sharma} R.,  {Jaleel} A.,   {Dutta} A.,  2017, \mn@doi
  [\mnras] {10.1093/mnrasl/slx039}, \href
  {http://adsabs.harvard.edu/abs/2017MNRAS.468L.118J} {468, L118}

\bibitem[\protect\citeauthoryear{{Kubota}, {Tanaka}, {Makishima}, {Ueda},
  {Dotani}, {Inoue}  \& {Yamaoka}}{{Kubota} et~al.}{1998}]{Kubota}
{Kubota} A.,  {Tanaka} Y.,  {Makishima} K.,  {Ueda} Y.,  {Dotani} T.,  {Inoue}
  H.,   {Yamaoka} K.,  1998, \mn@doi [\pasj] {10.1093/pasj/50.6.667}, \href
  {http://adsabs.harvard.edu/abs/1998PASJ...50..667K} {50, 667}

\bibitem[\protect\citeauthoryear{{Lewin} \& {van der Klis}}{{Lewin} \& {van der
  Klis}}{2006}]{2006book}
{Lewin} W.~H.~G.,  {van der Klis} M.,  2006, {Compact Stellar X-ray Sources}.
Cambridge University Press

\bibitem[\protect\citeauthoryear{{Lewin}, {Hoffman}, {Doty}  \&
  {Liller}}{{Lewin} et~al.}{1976}]{Lewin1976}
{Lewin} W.~H.~G.,  {Hoffman} J.~A.,  {Doty} J.,   {Liller} W.,  1976, \iaucirc,
  \href {http://adsabs.harvard.edu/abs/1976IAUC.2994....2L} {2994}

\bibitem[\protect\citeauthoryear{{Lin}, {Remillard}  \& {Homan}}{{Lin}
  et~al.}{2007}]{Lin07}
{Lin} D.,  {Remillard} R.~A.,   {Homan} J.,  2007, \mn@doi [\apj]
  {10.1086/521181}, \href {http://adsabs.harvard.edu/abs/2007ApJ...667.1073L}
  {667, 1073}

\bibitem[\protect\citeauthoryear{{Lin}, {Remillard}  \& {Homan}}{{Lin}
  et~al.}{2009}]{Lin09}
{Lin} D.,  {Remillard} R.~A.,   {Homan} J.,  2009, \mn@doi [\apj]
  {10.1088/0004-637X/696/2/1257}, \href
  {http://adsabs.harvard.edu/abs/2009ApJ...696.1257L} {696, 1257}

\bibitem[\protect\citeauthoryear{{Mondal}, {Dewangan}, {Pahari}, {Misra},
  {Kembhavi}  \& {Raychaudhuri}}{{Mondal} et~al.}{2016}]{Mondal2016}
{Mondal} A.~S.,  {Dewangan} G.~C.,  {Pahari} M.,  {Misra} R.,  {Kembhavi}
  A.~K.,   {Raychaudhuri} B.,  2016, \mn@doi [\mnras] {10.1093/mnras/stw1464},
  \href {http://adsabs.harvard.edu/abs/2016MNRAS.461.1917M} {461, 1917}

\bibitem[\protect\citeauthoryear{{Mu{\~n}oz-Darias}, {Fender}, {Motta}  \&
  {Belloni}}{{Mu{\~n}oz-Darias} et~al.}{2014}]{Munoz}
{Mu{\~n}oz-Darias} T.,  {Fender} R.~P.,  {Motta} S.~E.,   {Belloni} T.~M.,
  2014, \mn@doi [\mnras] {10.1093/mnras/stu1334}, \href
  {http://adsabs.harvard.edu/abs/2014MNRAS.443.3270M} {443, 3270}

\bibitem[\protect\citeauthoryear{{Muno}, {Chakrabarty}, {Galloway}  \&
  {Savov}}{{Muno} et~al.}{2001}]{Muno}
{Muno} M.~P.,  {Chakrabarty} D.,  {Galloway} D.~K.,   {Savov} P.,  2001,
  \mn@doi [\apjl] {10.1086/320682}, \href
  {http://adsabs.harvard.edu/abs/2001ApJ...553L.157M} {553, L157}

\bibitem[\protect\citeauthoryear{{Negoro} et~al.,}{{Negoro}
  et~al.}{2015}]{Negoro}
{Negoro} H.,  et~al., 2015, The Astronomer's Telegram, \href
  {http://adsabs.harvard.edu/abs/2015ATel.7943....1N} {7943}

\bibitem[\protect\citeauthoryear{{Neilsen} \& {Lee}}{{Neilsen} \&
  {Lee}}{2009}]{Neilsen2009}
{Neilsen} J.,  {Lee} J.~C.,  2009, \mn@doi [\nat] {10.1038/nature07680}, \href
  {http://adsabs.harvard.edu/abs/2009Natur.458..481N} {458, 481}

\bibitem[\protect\citeauthoryear{{Ono}, {Makishima}, {Sakurai}, {Zhang},
  {Yamaoka}  \& {Nakazawa}}{{Ono} et~al.}{2017}]{Ono2017}
{Ono} K.,  {Makishima} K.,  {Sakurai} S.,  {Zhang} Z.,  {Yamaoka} K.,
  {Nakazawa} K.,  2017, \mn@doi [\pasj] {10.1093/pasj/psw126}, \href
  {http://adsabs.harvard.edu/abs/2017PASJ...69...23O} {69, 23}

\bibitem[\protect\citeauthoryear{{Oosterbroek}, {Parmar}, {Sidoli}, {in 't
  Zand}  \& {Heise}}{{Oosterbroek} et~al.}{2001}]{Oosterbroek}
{Oosterbroek} T.,  {Parmar} A.~N.,  {Sidoli} L.,  {in 't Zand} J.~J.~M.,
  {Heise} J.,  2001, \mn@doi [\aap] {10.1051/0004-6361:20011006}, \href
  {http://adsabs.harvard.edu/abs/2001A%26A...376..532O} {376, 532}

\bibitem[\protect\citeauthoryear{{Parikh}, {Wijnands}, {Bahramian}, {Degenaar}
  \& {Heinke}}{{Parikh} et~al.}{2017}]{Parikh}
{Parikh} A.,  {Wijnands} R.,  {Bahramian} A.,  {Degenaar} N.,   {Heinke} C.,
  2017, The Astronomer's Telegram, \href
  {http://adsabs.harvard.edu/abs/2017ATel10169....1P} {10169}

\bibitem[\protect\citeauthoryear{{Ponti}, {Fender}, {Begelman}, {Dunn},
  {Neilsen}  \& {Coriat}}{{Ponti} et~al.}{2012}]{Ponti12}
{Ponti} G.,  {Fender} R.~P.,  {Begelman} M.~C.,  {Dunn} R.~J.~H.,  {Neilsen}
  J.,   {Coriat} M.,  2012, \mn@doi [\mnras]
  {10.1111/j.1745-3933.2012.01224.x}, \href
  {http://adsabs.harvard.edu/abs/2012MNRAS.422L..11P} {422, L11}

\bibitem[\protect\citeauthoryear{{Ponti}, {Mu{\~n}oz-Darias}  \&
  {Fender}}{{Ponti} et~al.}{2014}]{Ponti14}
{Ponti} G.,  {Mu{\~n}oz-Darias} T.,   {Fender} R.~P.,  2014, \mn@doi [\mnras]
  {10.1093/mnras/stu1742}, \href
  {http://adsabs.harvard.edu/abs/2014MNRAS.444.1829P} {444, 1829}

\bibitem[\protect\citeauthoryear{{Ponti} et~al.,}{{Ponti}
  et~al.}{2015}]{PontiAX-2015}
{Ponti} G.,  et~al., 2015, \mn@doi [\mnras] {10.1093/mnras/stu1853}, \href
  {http://adsabs.harvard.edu/abs/2015MNRAS.446.1536P} {446, 1536}

\bibitem[\protect\citeauthoryear{{Ponti}, {Bianchi}, {Mu{\~n}oz-Darias}, {De},
  {Fender}  \& {Merloni}}{{Ponti} et~al.}{2016}]{Ponti2016}
{Ponti} G.,  {Bianchi} S.,  {Mu{\~n}oz-Darias} T.,  {De} K.,  {Fender} R.,
  {Merloni} A.,  2016, \mn@doi [Astronomische Nachrichten]
  {10.1002/asna.201612339}, \href
  {http://adsabs.harvard.edu/abs/2016AN....337..512P} {337, 512}

\bibitem[\protect\citeauthoryear{{Ponti} et~al.,}{{Ponti}
  et~al.}{2018}]{PontiAX}
{Ponti} G.,  et~al., 2018, \mn@doi [\mnras] {10.1093/mnras/stx2425}, \href
  {http://adsabs.harvard.edu/abs/2018MNRAS.473.2304P} {473, 2304}

\bibitem[\protect\citeauthoryear{{Raichur}, {Misra}  \& {Dewangan}}{{Raichur}
  et~al.}{2011}]{Raichur2011}
{Raichur} H.,  {Misra} R.,   {Dewangan} G.,  2011, \mn@doi [\mnras]
  {10.1111/j.1365-2966.2011.19075.x}, \href
  {http://adsabs.harvard.edu/abs/2011MNRAS.416..637R} {416, 637}

\bibitem[\protect\citeauthoryear{{Raman}, {Maitra}  \& {Paul}}{{Raman}
  et~al.}{2018}]{Raman2018}
{Raman} G.,  {Maitra} C.,   {Paul} B.,  2018, \mn@doi [\mnras]
  {10.1093/mnras/sty918}, \href
  {http://adsabs.harvard.edu/abs/2018MNRAS.477.5358R} {477, 5358}

\bibitem[\protect\citeauthoryear{{Reeves}, {Done}, {Pounds}, {Terashima},
  {Hayashida}, {Anabuki}, {Uchino}  \& {Turner}}{{Reeves}
  et~al.}{2008}]{Reeves}
{Reeves} J.,  {Done} C.,  {Pounds} K.,  {Terashima} Y.,  {Hayashida} K.,
  {Anabuki} N.,  {Uchino} M.,   {Turner} M.,  2008, \mn@doi [\mnras]
  {10.1111/j.1745-3933.2008.00443.x}, \href
  {http://adsabs.harvard.edu/abs/2008MNRAS.385L.108R} {385, L108}

\bibitem[\protect\citeauthoryear{{Romano} et~al.,}{{Romano}
  et~al.}{2006}]{Romano}
{Romano} P.,  et~al., 2006, \mn@doi [\aap] {10.1051/0004-6361:20065071}, \href
  {http://ukads.nottingham.ac.uk/abs/2006A%26A...456..917R} {456, 917}

\bibitem[\protect\citeauthoryear{{Sakurai}, {Yamada}, {Torii}, {Noda},
  {Nakazawa}, {Makishima}  \& {Takahashi}}{{Sakurai}
  et~al.}{2012}]{Sakurai2012}
{Sakurai} S.,  {Yamada} S.,  {Torii} S.,  {Noda} H.,  {Nakazawa} K.,
  {Makishima} K.,   {Takahashi} H.,  2012, \mn@doi [\pasj]
  {10.1093/pasj/64.4.72}, \href
  {http://adsabs.harvard.edu/abs/2012PASJ...64...72S} {64, 72}

\bibitem[\protect\citeauthoryear{{Sharma}, {Jaleel}, {Jain}, {Paul}  \&
  {Dutta}}{{Sharma} et~al.}{2018}]{Sharma}
{Sharma} R.,  {Jaleel} A.,  {Jain} C.,  {Paul} B.,   {Dutta} A.,  2018, \mn@doi
  [Journal of Astrophysics and Astronomy] {10.1007/s12036-017-9497-y}, \href
  {http://adsabs.harvard.edu/abs/2018JApA...39...16S} {39, 16}

\bibitem[\protect\citeauthoryear{{Sidoli}, {Oosterbroek}, {Parmar}, {Lumb}  \&
  {Erd}}{{Sidoli} et~al.}{2001}]{Sidoli}
{Sidoli} L.,  {Oosterbroek} T.,  {Parmar} A.~N.,  {Lumb} D.,   {Erd} C.,  2001,
  \mn@doi [\aap] {10.1051/0004-6361:20011322}, \href
  {http://adsabs.harvard.edu/abs/2001A%26A...379..540S} {379, 540}

\bibitem[\protect\citeauthoryear{{Sidoli}, {Parmar}, {Oosterbroek}  \&
  {Lumb}}{{Sidoli} et~al.}{2002}]{Sidoli2002}
{Sidoli} L.,  {Parmar} A.~N.,  {Oosterbroek} T.,   {Lumb} D.,  2002, \mn@doi
  [\aap] {10.1051/0004-6361:20020192}, \href
  {http://adsabs.harvard.edu/abs/2002A%26A...385..940S} {385, 940}

\bibitem[\protect\citeauthoryear{{Titarchuk}}{{Titarchuk}}{1994}]{Titarchuk}
{Titarchuk} L.,  1994, \mn@doi [\apj] {10.1086/174760}, \href
  {http://adsabs.harvard.edu/abs/1994ApJ...434..570T} {434, 570}

\bibitem[\protect\citeauthoryear{{Verner}, {Ferland}, {Korista}  \&
  {Yakovlev}}{{Verner} et~al.}{1996}]{Verner}
{Verner} D.~A.,  {Ferland} G.~J.,  {Korista} K.~T.,   {Yakovlev} D.~G.,  1996,
  \mn@doi [\apj] {10.1086/177435}, \href
  {http://adsabs.harvard.edu/abs/1996ApJ...465..487V} {465, 487}

\bibitem[\protect\citeauthoryear{{Wijnands}, {Strohmayer}  \&
  {Franco}}{{Wijnands} et~al.}{2001}]{WijnandsBO}
{Wijnands} R.,  {Strohmayer} T.,   {Franco} L.~M.,  2001, \mn@doi [\apjl]
  {10.1086/319128}, \href {http://adsabs.harvard.edu/abs/2001ApJ...549L..71W}
  {549, L71}

\bibitem[\protect\citeauthoryear{{Wijnands}, {Muno}, {Miller}, {Franco},
  {Strohmayer}, {Galloway}  \& {Chakrabarty}}{{Wijnands}
  et~al.}{2002}]{Wijnandsbb}
{Wijnands} R.,  {Muno} M.~P.,  {Miller} J.~M.,  {Franco} L.~M.,  {Strohmayer}
  T.,  {Galloway} D.,   {Chakrabarty} D.,  2002, \mn@doi [\apj]
  {10.1086/338140}, \href {http://adsabs.harvard.edu/abs/2002ApJ...566.1060W}
  {566, 1060}

\bibitem[\protect\citeauthoryear{{Wilms}, {Allen}  \& {McCray}}{{Wilms}
  et~al.}{2000}]{Wilms}
{Wilms} J.,  {Allen} A.,   {McCray} R.,  2000, \mn@doi [\apj] {10.1086/317016},
  \href {http://adsabs.harvard.edu/abs/2000ApJ...542..914W} {542, 914}

\bibitem[\protect\citeauthoryear{{Yamaoka}, {Ueda}, {Inoue}, {Nagase},
  {Ebisawa}, {Kotani}, {Tanaka}  \& {Zhang}}{{Yamaoka} et~al.}{2001}]{Yamaoka}
{Yamaoka} K.,  {Ueda} Y.,  {Inoue} H.,  {Nagase} F.,  {Ebisawa} K.,  {Kotani}
  T.,  {Tanaka} Y.,   {Zhang} S.~N.,  2001, \mn@doi [\pasj]
  {10.1093/pasj/53.2.179}, \href
  {http://adsabs.harvard.edu/abs/2001PASJ...53..179Y} {53, 179}

\bibitem[\protect\citeauthoryear{{Zdziarski}, {Johnson}  \&
  {Magdziarz}}{{Zdziarski} et~al.}{1996}]{Zdziarski}
{Zdziarski} A.~A.,  {Johnson} W.~N.,   {Magdziarz} P.,  1996, \mn@doi [\mnras]
  {10.1093/mnras/283.1.193}, \href
  {http://adsabs.harvard.edu/abs/1996MNRAS.283..193Z} {283, 193}

\bibitem[\protect\citeauthoryear{{Zhang}, {Makishima}, {Sakurai}, {Sasano}  \&
  {Ono}}{{Zhang} et~al.}{2014}]{Zhang2014}
{Zhang} Z.,  {Makishima} K.,  {Sakurai} S.,  {Sasano} M.,   {Ono} K.,  2014,
  \mn@doi [\pasj] {10.1093/pasj/psu117}, \href
  {http://adsabs.harvard.edu/abs/2014PASJ...66..120Z} {66, 120}

\bibitem[\protect\citeauthoryear{{Zhang}, {Sakurai}, {Makishima}, {Nakazawa},
  {Ono}, {Yamada}  \& {Xu}}{{Zhang} et~al.}{2016}]{Zhang2016}
{Zhang} Z.,  {Sakurai} S.,  {Makishima} K.,  {Nakazawa} K.,  {Ono} K.,
  {Yamada} S.,   {Xu} H.,  2016, \mn@doi [\apj] {10.3847/0004-637X/823/2/131},
  \href {http://adsabs.harvard.edu/abs/2016ApJ...823..131Z} {823, 131}

\bibitem[\protect\citeauthoryear{{{\.Z}ycki}, {Done}  \& {Smith}}{{{\.Z}ycki}
  et~al.}{1999}]{Zycki}
{{\.Z}ycki} P.~T.,  {Done} C.,   {Smith} D.~A.,  1999, \mn@doi [\mnras]
  {10.1046/j.1365-8711.1999.02885.x}, \href
  {http://adsabs.harvard.edu/abs/1999MNRAS.309..561Z} {309, 561}

\bibitem[\protect\citeauthoryear{in~'t Zand et~al.,}{in~'t Zand
  et~al.}{1999a}]{intZand1999}
in~'t Zand J.~J.~M.,  et~al., 1999a, \aap, \href
  {http://adsabs.harvard.edu/abs/1999A%26A...345..100I} {345, 100}

\bibitem[\protect\citeauthoryear{in~'t Zand, {Heise}, {Smith}, {Cocchi},
  {Natalucci}, {Celidonio}, {Augusteijn}  \& {Freyhammer}}{in~'t Zand
  et~al.}{1999b}]{intZand}
in~'t Zand J.,  {Heise} J.,  {Smith} M.~J.~S.,  {Cocchi} M.,  {Natalucci} L.,
  {Celidonio} G.,  {Augusteijn} T.,   {Freyhammer} L.,  1999b, \iaucirc, \href
  {http://adsabs.harvard.edu/abs/1999IAUC.7138....1I} {7138}

\bibitem[\protect\citeauthoryear{van~der Klis}{van~der Klis}{2006}]{vanderKlis}
van~der Klis M.,  2006, in {Lewin} W.,  {van der Klis} M.,  eds, Compact
  Stellar X-ray Sources. Cambridge University Press.
p.~39 (\mn@eprint {arXiv} {astro-ph/0410551})

\makeatother
\end{thebibliography}






\label{lastpage}
\end{document}